\documentclass[12pt]{article}
\usepackage[english]{babel}

\usepackage{amssymb,amsmath}

\usepackage{amsfonts}
\usepackage{latexsym}
\usepackage{graphicx,color}
\usepackage{epsfig}
\usepackage{verbatim}
\usepackage{amsthm}
\usepackage{amsbsy}

\usepackage{hyperref}
\usepackage{cite}
\definecolor{link}{rgb}{.8,.15,.1}
\hypersetup{colorlinks=true,linkcolor=link,citecolor=link,urlcolor=link,linktocpage}
\newcommand{\be}{\begin{equation}}
\newcommand{\ee}{\end{equation}}
\newcommand{\bi}{\begin{itemize}}
\newcommand{\ei}{\end{itemize}}
\newcommand{\bea}{\begin{eqnarray}}
\newcommand{\eea}{\end{eqnarray}}
\newcommand{\ba}{\begin{array}}
\newcommand{\ea}{\end{array}}

\def\ym2{{YM$_2$}}

\def\Tr{\mathrm{Tr}}
\def\rmd{{\rm d}}
\def\rmD{{\rm D}}

\newcommand{\newsection}[1]{\section{#1}\setcounter{equation}{0}}
\newcommand{\nn}{\nonumber}
\begin{document}

\begin{titlepage}

\rightline{KIAS-P16082}
\vskip 1cm

\vskip 1cm
\begin{center}
\centerline{\large \bf The coupling of Poisson sigma models to topological backgrounds}
\vspace{1cm}

Dario Rosa \\

\bigskip
  School of Physics, Korea Institute for Advanced Study, Seoul 02455, Korea\\
\smallskip
Dario85@kias.re.kr


\vskip .5in
{\bf Abstract }

\end{center}

We extend the coupling to the topological backgrounds, recently worked out for the 2-dimensional BF-model, to the most general Poisson sigma models. The coupling involves the choice of a Casimir function on the target manifold and modifies the BRST transformations. This in turn induces a change in the BRST cohomology of the resulting theory. The observables of the coupled theory are analyzed and their geometrical interpretation is given. We finally couple the theory to 2-dimensional topological gravity: this is the first step to study a topological string theory in propagation on a Poisson manifold. As an application, we show that the gauge-fixed vectorial supersymmetry of the Poisson sigma models has a natural explanation in terms of the theory coupled to topological gravity.

\vskip .1in

\thispagestyle{empty}

\vfill
\eject

\end{titlepage}

\setcounter{footnote}{0}

\tableofcontents

\newsection{Introduction}
\label{sec:intro}

Topological field theories (TFTs) are often divided in two classes:  Schwarz's TFTs and Witten's TFTs. The peculiar feature of the former is that their classical action is manifestly independent from the metric on the manifold on which the theory is defined.\footnote{Of course, the gauge-fixing term is not  metric-independent.} Standard examples of such theories are the 3-dimensional Chern-Simons theory and the BF-models. On the other hand, Witten's TFTs are usually obtained starting from a supersymmetric field theory and then performing a topological twist: after the twist one of the original, spinorial, supercharges turns out to be a scalar and it can be treated as a BRST operator for the resulting theory. Such a BRST symmetry (that one can call topological supersymmetry, to remember its origin from a supersymmetric field theory) is so large that {\it all} the local propagating degrees of freedom are BRST-exact and therefore they do not affect the correlators. On the other hand, Schwarz's TFTs do not enjoy usually any topological supersymmetry.

What Schwarz's TFTs and Witten's TFTs have in common is that, in both cases,  all the dynamics is encoded in the global, non propagating, degrees of freedom. From this point of view it is natural to think that these two types of theories could be  two different faces of the same coin and that could be related in a closer way. Indeed, for some particular cases, such a relation can be found: for example, it has been noticed \cite{Bonechi:2007ar}, \cite{Bonechi:2016wqz} that the topological A-model (a Witten's TFT) can be obtained by performing a suitable gauge-fixing of a particular Poisson sigma model (PSM), which is a Schwarz's TFT that we will describe in details later.

Starting from \cite{Imbimbo:2009dy} for the case of 3-dimensional Chern-Simons, and then in subsequent papers \cite{Giusto:2012jm}, \cite{Imbimbo:2014pla} and \cite{Bae:2015eoa}, a completely different connection between Witten's TFTs and Schwarz's TFTs started to emerge: by coupling a Schwarz's TFT  to some additional topological {\it background} multiplets (including topological gravity plus some other topological multiplets which depend on the context) a topological supersymmetry appears.  Since, as we recalled, the topological supersymmetry is the prominent feature of a Witten's TFT, one understands that the coupling of a Schwarz's TFT to topological backgrounds produces a (family of) Witten's TFT(s). The original Schwarz's TFT has to be considered as a particular point in the space of the theories coupled to the topological backgrounds: in generic points of the space of  backgrounds the theory has the topological supersymmetry, but in some very specific points of the space of backgrounds the topological supersymmetry collapses and one ends up with the original Schwarz's TFT.

From a technical point of view, the coupling to the topological backgrounds produces a BRST operator which  usually is nilpotent only on-shell. A standard method to treat BRST operators nilpotent only on-shell is the so-called BV formalism.\footnote{See, for example, \cite{Fuster:2005eg} for a concise introduction.} Let us recall its main features: for each field one includes a corresponding antifield. Then one modifies the BRST variations of the fields (and defines the BRST variations of the antifields) such that at the end the BRST operator is nilpotent off-shell. However the antifields are {\it not} treated as independent fields: during the gauge-fixing procedure they are fixed to some functionals of the fields. 

On the other hand, it has been observed in \cite{Imbimbo:2014pla} and \cite{Bae:2015eoa} that the  BRST algebra including also the antifields is equivalent to a {\it twisted} version of the SUSY algebra of a vector multiplet in a corresponding supersymmetric field theory. This observation has led to a new interesting application of the BV algorithm: instead of treating the antifields as functionals of the fields (and viewing the BV formalism as a way to gauge-fix a BRST symmetry which closes only on-shell), the antifields can be treated as {\it independent} auxiliary fields, whose role is just to ensure the closure of the algebra off-shell (this is indeed the role that auxiliary fields play in supersymmetric field theories). Of course, it must be emphasized that treating the antifields as auxiliary fields gives a theory which, a priori, is {\it different} from the theory obtained via the traditional BV method: what we are saying here is just that the BV algorithm provides an efficient way to construct theories with a topological supersymmetry that closes off-shell. Following this recipe one finds a close and new relation between topological field theories (coupled to the topological backgrounds) and supersymmetric field theories on curved space. This correspondence allows to understand some of the results obtained via supersymmetric localization in the last ten years (see \cite{Pestun:2016zxk} for an exhaustive review) from a topological and cohomological point of view, and in many cases (discussed in \cite{Imbimbo:2014pla} and \cite{Bae:2015eoa}) such a cohomological viewpoint has provided an extension of the results obtained via standard supersymmetric field theories. 

In this paper we provides a generalization of the results obtained in \cite{Bae:2015eoa}: in that paper it is shown that the 2-dimensional BF-model can be consistently coupled to a topological $U(1)$ multiplet. After the coupling one obtains a topological description of the non-topological theory of 2-dimensional YM. Moreover, the theory coupled to the background multiplet acquires a topological supersymmetry that in the standard treatment of 2-dimensional YM is introduced ``by hand''. This gives a very explicit example of the phenomenon we mentioned before: after the coupling to the topological background, a topological supersymmetry emerges.

The 2-dimensional BF-model is a particular example of a set of  TFTs  which go under the name of Poisson sigma models (PSMs) \cite{Schaller:1994es}: they are topological sigma models in which the target space is a Poisson manifold. Let us recall that a Poisson manifold is a manifold $M$ provided with a bivector field, $\Pi^{ij} \in \Gamma(\wedge^2 TM)$, which satisfies the condition
\bea
&& \left[\Pi \, , \, \Pi \right]_S \equiv \Pi^{il} \partial_l \Pi^{jk} + \Pi^{jl} \partial_l \Pi^{ki} + \Pi^{kl} \partial_l \Pi^{ij} = 0 \ ,
\eea
known as Jacobi condition. The BF-model can be seen as a particular case in which the target space is a {\it linear} Poisson manifold, i.e. a Poisson manifold whose corresponding bivector $\Pi^{ij}$ is linear in the local coordinates.

In this paper we will show that the coupling to the topological $U(1)$ multiplet, worked out in \cite{Bae:2015eoa} for a BF-model, can be generalized to include all the possible PSMs. The procedure to obtain the coupled theory is  similar, but more involved, to the one used for the BF-model. The coupling to the $U(1)$ background multiplet is done by choosing a {\it Casimir function} $C(X)$ on $M$ (which, as we will review, is a function  invariant under the action of the Poisson bivector $\Pi^{ij}$). After the coupling, both the BRST variations and the action are modified by terms involving the Casimir function $C(X)$, and this changes the observables of the theory: before the coupling to the topological $U(1)$ multiplet, the observables can be identified with the elements of the {\it Poisson cohomology} of $M$; after the coupling they are identified with the elements of the Poisson cohomology of $M$ which { also} have vanishing {\it Schouten bracket} with the Casimir function $C(X)$. We will also see that the coupled system admits new composite observables constructed out of the PSM fields and the $U(1)$ background fields. 

The resulting model is further coupled to 2-dimensional topological gravity: this is the first step to construct a topological string theory in propagation on the target space $M$, an important  problem that we hope to address in the future and that could lead to define enumerative invariants for a Poisson manifold very similar to the GW invariants. The observables of the system coupled to topological gravity are discussed.

Let us also mention that the system of backgrounds (the topological $U(1)$ multiplet and $2$-dimensional topological gravity) that we considered in this paper is equivalent to $2$-dimensional $N=(2,2)$ supergravity \cite{Bae:2015eoa}, therefore it is conceivable that the study of the PSMs coupled to these backgrounds could provide a way to define new $2$-dimensional supersymmetric vector multiplets. This is another important aspect that we hope to address in the future. 

The paper is organized as follow. In Section \ref{sec:2dYM}, to make the paper self-contained, we review the coupling of the 2-dimensional BF-model to the topological $U(1)$ background multiplet. We also use this Section to explain the new application of the BV technique we described in this Introduction. This leads naturally to the topological supersymmetry. The role of the ``gaugino'' field, usually introduced by hand in 2-dimensional YM, is here played by a combination between $\gamma^{(0)}$, the ghost-for-ghost of the topological $U(1)$ background multiplet, and $A^\ast$, the antifield of the gauge field $A$. In Section \ref{sec:PSMs} we start by reviewing the standard construction of a generic PSM, following mostly \cite{Cattaneo:1999fm}, and then we see how the coupling to the topological $U(1)$ background is carried out for these more general models.\footnote{Let us also mention that, for a generic PSM, treating the antifields as indipendent fields gives a theory which is effectively different from the original PSM: this is a possibility that we already mentioned and indeed the PSM provides an explicit example of such a phenomenon. For this reason we also make a discussion about the coupling to the topological backgrounds in the BV formalism.} 
A discussion of the observables is given and we also see that the topological action is equivalent, in the relevant cohomology, to a purely algebraic observable. In Section \ref{sec:couplinggravity} we  couple the resulting system to topological gravity. The observables identified in Section \ref{sec:PSMs} are immediately promoted to observables of the theory coupled to topological gravity. As an application we discuss how the gauge-fixed vectorial supersymmetry discovered in \cite{Cattaneo:1999fm} can be easily understood in terms of the theory coupled to topological gravity. Finally,  Section \ref{sec:conclusions} contains the conclusions and some ideas for future works.

\newsection{A review of two-dimensional Yang-Mills coupled to topological backgrounds}
\label{sec:2dYM}

In this section, we review the coupling of 2-dimensional YM to topological backgrounds \cite{Bae:2015eoa}.  The generalization to generic PSMs will be discussed in the next Section. 
  
It is commonly said that 2-dimensional Yang-Mills is a topological theory since gauge invariance in two dimensions removes all the propagating local degrees of freedom. However, the 2-dimensional YM action is not topological, i.e. it is not independent from the 2-dimensional metric. It is convenient to write the 2-dimensional action in a slightly unusual way \cite{Witten:1992xu} by introducing, beyond the gauge field $A = A^a T^a$, an additional adjoint scalar $\phi = \phi^a T^a$ \footnote{Both $A$ and $\phi$ have ghost number $0$ and $T^a$, with $a = 1 \dots \mathrm{dim} G$, are the generators of the Lie algebra associated to the group $G$.} 
\bea\label{eq:YMactionintro}
&& \Gamma_{\mathrm{YM}} = \int_{\Sigma} \Tr \, (\phi \, F) +  \epsilon \int_\Sigma \rmd^2 x \sqrt{g} \, \frac 12 \, \Tr\,( \phi^2) \ ,
\eea
where $\epsilon$ is a constant proportional to the square of the standard YM coupling constant, $\Sigma$ is a 2-dimensional Riemann surface\footnote{Since the main goal of the paper is to explain the procedure of coupling the topological theory to topological backgrounds, we will restrict to the case in which $\Sigma$ is {\it closed}. The discussion can be generalized to the case in which $\Sigma$ has boundaries.} provided with a metric $g$ and $F$ is the field-strength two-form
\bea
F = \rmd \, A + A^2.
\eea
The correspondence with the standard YM action is then recovered by integrating out $\phi$
\bea
&& \Gamma^\prime_{\mathrm{YM}} = - \frac{1}{ \epsilon} \int_{\Sigma}\rmd^2 x \sqrt{g} \, \frac 12 \, \Tr \,(F^2) \ .
\eea
On the other hand, (\ref{eq:YMactionintro}) makes explicit the dependence of the theory from the 2-dimensional metric, via the volume form $\rmd^2 x\sqrt g$ which appears in the term
\bea\label{eq:nontoptermintro}
&& \frac \epsilon 2 \int_\Sigma \rmd^2 x \sqrt{g} \, \Tr \, (\phi^2) \ .
\eea
The action (\ref{eq:YMactionintro}) also shows that, at least classically, the dependence from the metric gets removed by considering the $\epsilon \to 0$ limit: in this way one obtains the topological action
\bea\label{eq:topYMactionintro}
 && \Gamma_{\mathrm{YM}} |_{\epsilon = 0} = \int_\Sigma \Tr \, (\phi F) \ ,
\eea
which is manifestly independent from the metric. It is indeed the action for a 2-dimensional BF-model: a topological theory of Schwarz's type.

Both the physical action (\ref{eq:YMactionintro}) and the topological action (\ref{eq:topYMactionintro}), do {\it not} possess any topological supersymmetry: they are only invariant under gauge BRST transformations
\bea\label{eq:sgaugeintro}
 	&& s_{gauge} c = - c^2 \ , \nonumber \\
	&& s_{gauge} A = - \rmD \, c \ , \nonumber \\
	&& s_{gauge} \phi = - [c, \phi] \ , 
\eea
where, as usual, $c = c^a T^a$ is the ghost field associated to the gauge invariance and the gauge covariant derivative is\footnote{In this paper we will adopt the convention that the BRST operator and the external differential {\it anticommute}.}
\bea
&& \rmD \, c \equiv \rmd \, c + [A,c]_+ \ ,
\eea
whereas it is not present any topological supersymmetry, like the one of 4-dimensional topological YM
\bea
\label{eq:psisymmetry}
	s \, A = \psi \ ,
\eea
where $\psi$ is the topological gaugino (a 1-form of ghost number $1$). The lacking of a topological supersymmetry is not surprising, since 2-dimensional YM is a deformation of a Schwarz's TFT.

However  the action $\Gamma_{YM}\big|_{\epsilon=0}$ can be easily supersymmetrized, by adding a decoupled quadratic fermionic term \cite{Witten:1992xu}
\bea
\label{eq:Wittenaction}
\Gamma_{top} = \Gamma_{YM}\big|_{\epsilon=0}-\frac{1}{2}\int_{\Sigma}  \,\Tr\, \psi\wedge\psi =\int_{\Sigma} \Tr \,\phi \,F - \frac{1}{2}\int_{\Sigma} \,\Tr\, \psi\wedge\psi \ .
\label{ActionTopological}
\eea
This action is indeed invariant under topological Yang-Mills BRST transformations\footnote{We will use the symbol $s_0$ to denote the {\it rigid} BRST operator, to be distinguished from the BRST operator coupled to topological gravity which we will denote with $s$ and that will be discussed in Section \ref{sec:couplinggravity}.}
\bea
\label{eq:susytransformationYM}
&& s_0\, c=  - c^2 + \phi \ ,\nonumber\\
&& s_0\, A = - \rmD\, c +\psi \ , \nonumber\\
&& s_0\, \psi = -[c, \psi]- \rmD\,\phi \ ,\nonumber\\
&& s_0\, \phi = -[c, \phi] \ .
\label{BRSTtopologicalYM}
\eea
Notice that the deformation (\ref{eq:susytransformationYM}) shifts the ghost number of the scalar field $\phi$: it has now ghost number $+2$. This shift makes harder to generalize the deformation (\ref{eq:Wittenaction}) to more general PSMs. We will see that the coupling to topological backgrounds gives a way to overcome this difficulty.

By switching on $\epsilon$, one obtains the final action
\bea
\Gamma_{W} = \int_{\Sigma} \Tr \,\phi \,F  + \frac{\epsilon}{2} \int_{\Sigma} \rmd^2x\,\sqrt{g}\, \Tr\, \phi^2 \ - \frac{1}{2}\int_{\Sigma} \,\Tr\, \psi\wedge\psi .
\label{ActionWitten}
\eea
$\Gamma_{W}$ is also invariant under  (\ref{BRSTtopologicalYM}); nevertheless it  is not fully topological,  since it explicitly depends on a 2-dimensional background metric via the volume form $\rmd^2x\, \sqrt{g}$. 

The construction of \cite{Bae:2015eoa}, that we are going to review, shows that {\it both} the non-topological deformation (\ref{eq:nontoptermintro}) and the fermionic term
\bea
\label{eq:psiterm}
\frac{1}{2}\int_{\Sigma}  \,\Tr\, \psi\wedge\psi \ ,
\eea
can be obtained by considering the topological BF-theory (\ref{eq:topYMactionintro}) coupled to an abelian topological {\it background} $U(1)$ multiplet. In this way one obtains a {\it topological} formulation of 2-dimensional YM and a more natural understanding of the topological supersymmetry transformations (\ref{eq:susytransformationYM}).

The main idea of \cite{Bae:2015eoa} is to replace both the metric and the coupling constant $\epsilon$  with a topological background and to extend the BRST action on the background.
This produces automatically the term (\ref{eq:psiterm}), introduced by hand  in  \cite{Witten:1992xu}. 

Let then  $f^{(2)}$ be a 2-form field and let us replace  the action (\ref{eq:YMactionintro}) with
\bea
\Gamma_1=   \int_{\Sigma} \Tr \,\phi \,F - \frac{1}{2} \int_{\Sigma} f^{(2)}\Tr\, \phi^2 \, \  .
\label{actionYMone}
\eea
This action is not equivalent to the original one.  A generic
$f^{(2)}$ admits a Hodge decomposition
\bea
f^{(2)} = \Omega^{(2)} + \rmd\, \Omega^{(1)} \ ,
\eea
where 
\bea
\Omega^{(2)} = \epsilon\, \rmd^2x \, \sqrt{g} \ ,
\eea
is a  representative of $H^{2}(\Sigma)$ and $\Omega^{(1)}$ a  1-form.  
For  $\Gamma_1$ to be equivalent to $\Gamma_{YM}$  we must remove the degrees of freedom associated
to $\Omega^{(1)}$. We do this by introducing a BRST symmetry for the background $f^{(2)}$ \footnote{The idea of extending the BRST symmetry to {\it physical} coupling constants has been introduced, in a different context, in \cite{Giusto:2012jm}.}
\bea
s_0\, f^{(2)} = - \rmd\, \psi^{(1)} \ ,
\label{BRSTone}
\eea
where $\psi^{(1)}$ is a fermionic background 1-form field of ghost number +1. The BRST transformation (\ref{BRSTone}) is degenerate: therefore we introduce also a scalar ghost-for-ghost background field $\gamma^{(0)}$ of ghost number +2 
\bea
&& s_0\, \psi^{(1)} = - \rmd\, \gamma^{(0)} \ ,
\label{BRSTtwo}
\eea
with
\bea
&& \qquad s_0\,\gamma^{(0)}=0 \ .
\label{BRSTthree}
\eea
However BRST-invariance is lost, since
\bea
&& s_0 \, \Gamma_1 =-s_0\,\bigl( \frac{1}{2} \int_{\Sigma} f^{(2)}\Tr\, \phi^2 \bigr)= \frac{1}{2} \int_{\Sigma} \rmd\, \psi^{(1)}\,\Tr\, \phi^2 =  - \int_{\Sigma} \psi^{(1)} \wedge \Tr (\phi \,\rmD\,\phi) \ .
\eea
To cure for this we modify the BRST transformation law for $A$
\bea
s_0\, A= - \rmD\,c   + \psi^{(1)}\,\phi +\cdots \ ,
\label{BRSTAdeformedone}
\eea
so that the BRST variation of the first term in $\Gamma_1$ cancels the lack of invariance of the second term:
\bea
s_0 \, \Gamma_1=  0 \ .
\eea
The problem with  (\ref{BRSTAdeformedone})  is that it is not nilpotent:
\bea
s^2_0\, A = - \rmd\,\gamma^{(0)}\,\phi+\cdots \ ,
\eea
to fix this it is necessary to deform the BRST transformation rule for the ghost $c$
\bea
s_0\, c = - c^2+ \gamma^{(0)}\, \phi \ .
\eea
With this modification one has
\bea
s^2_0\,c =0 \ ,
\eea
and also this induces an extra term in $s_0^2\,A$  which cancels the term proportional to $\rmd\,\gamma_0$:
\bea
s^2_0\, A = \rmD\, \bigl(\gamma^{(0)}\,\phi\bigr) - \rmd\,\gamma^{(0)}\,\phi +\cdots = \gamma^{(0)}\, \rmD\,\phi +\cdots \ .
\eea
Although this is still not zero, the lack of nilpotency is now reduced to a term proportional to the equations of motion of $A$:
\bea
\frac{\delta \Gamma_1}{\delta A} = \rmD\,\phi = 0 \ ,
\eea
and therefore
\bea
s^2_0 \, A = 0 \qquad \textrm{on\; shell} \ .
\eea
The BV formalism provides a systematic way to go off-shell. One introduces the antifield corresponding to $A$ 
 \be
A^{\ast}\equiv A^{\ast a}_\mu\, T^a\, dx^\mu \ ,
\ee
which is  a 1-form in the adjoint of the gauge group of ghost number -1,  and  an antifield dependent
term in the BRST transformation of $A$
\bea
&& s_0\, A = - \rmD\,c  + \psi^{(1)}\,\phi +\gamma^{(0)} \,A^\ast \ .
\label{BRSTA}
\eea
This makes $s_0$ nilpotent off-shell on all fields
\bea
s_0^2\, c= s_0^2\, A= s_0^2\,\phi=s_0^2\, A^\ast=0 \qquad \textrm{off\; shell} \ ,
\eea
as long as $A^\ast$  transforms according to
\bea
&& s_0\, A^\ast= - [c,  A^\ast] - \rmD\, \phi \ .
\eea
The new term proportional to $\gamma^{(0)}$ in (\ref{BRSTA})  spoils the invariance of the action
\bea
s_0\, \Gamma_1 =  - \int_{\Sigma} \Tr \,\phi\, \rmD\,\bigl(\gamma^{(0)}\,A^\ast)= \int_{\Sigma} \gamma^{(0)}\,\Tr \,\rmD\,\phi\wedge A^\ast \ ,
\eea
and this is anticipated in the BV framework: once an antifield dependent term is introduced in the BRST transformation of a field,
terms quadratic in the antifields must be added to the action.  Indeed the final, {\it topological}, action
\bea
\Gamma=   \int_{\Sigma} \Tr \,\phi \,F - \frac{1}{2} \int_{\Sigma} f^{(2)}\Tr\, \phi^2 +\frac{1}{2} \int_{\Sigma} \, \gamma^{(0)}\,\Tr\,  A^\ast\wedge A^\ast \ ,
\label{Actionoffshell}
\eea
is invariant:
\bea
s_0\, \Gamma =0 \ ,
\eea
under  BRST transformations of {\it both} fields and backgrounds\footnote{It can be observed that the action (\ref{Actionoffshell}) and the transformations (\ref{BRSTfieldsbackgrounds}) can be also understood via the so-called AKSZ formalism\cite{Alexandrov:1995kv}: they can be thought as obtained by considering two different BF-models (one of them abelian) and then coupling them via a cubic term. The author thanks A.~S.~Cattaneo for discussions on this point.}
\bea
&& s_0\, c = - c^2+ \gamma^{(0)}\, \phi \ ,\nonumber\\
&& s_0\, A = - \rmD\,c +\gamma^{(0)} \,A^\ast + \psi^{(1)}\, \phi \ ,\nonumber\\
&& s_0\, \phi = -[c, \phi] \ ,\nonumber\\
&& s_0\, A^\ast= - [c, A^\ast] - \rmD\, \phi \ ,\nonumber \\
&& s_0\, f^{(2)} = -\rmd\, \psi^{(1)} \ ,\nonumber\\
&& s_0\, \psi^{(1)} = - \rmd\, \gamma^{(0)} \ ,\nonumber\\
&& s_0\, \gamma^{(0)}=0 \ .
\label{BRSTfieldsbackgrounds}
\eea

\subsection{The topological supersymmetry}
\label{sec:topologicalSUSY}

We have seen that  the field $A^\ast$ emerges  naturally in the context of the BV formalism.
In the BV framework, the action  (\ref{Actionoffshell}) would not however be the full action. The
BV action is given by adding to (\ref{Actionoffshell}) a {\it canonical} piece, which schematically reads
\bea
\label{eq:canonicalscheme}
&& \Gamma_{\mathrm{can}}= \sum_{\Phi} \int_{\Sigma}  (s_0 \, \Phi) \, \Phi^{\ast} \ ,
\label{canonicalActionBV}
\eea
where we use the symbol $\Phi$ to collectively indicate all the fields and backgrounds.
The full BV action 
\bea
&&\Gamma_{\rm BV} = \Gamma + \Gamma_{\mathrm{can}} \ ,
\eea
generates the BRST transformations of both fields and antifields via the familiar BV formulas.

However, from an algebraic point of view, the interpretation of $A^\ast$ as the antifield of $A$ is not mandatory.  In \cite{Imbimbo:2014pla} and \cite{Bae:2015eoa} it has been observed that an alternative --- although exotic ---  interpretation is available and it leads naturally
to the topological supersymmetry (\ref{eq:susytransformationYM}) and to the action (\ref{eq:Wittenaction}). Let us review this interpretation.

In the new approach (that we could call supersymmetric to recall that it realizes a connection with the study of supersymmetric field theories on curved spaces) $A^{\ast}$  is seen as  an {\it independent} auxiliary field, whose role is to close the BRST transformations off-shell: at the same time,  the action is taken to be $\Gamma$,  disregarding the canonical piece $\Gamma_{\mathrm{can}}$.

This approach is consistent since the BRST transformations close on the fields $(\phi, A, c, A^\ast)$ and leave  $\Gamma$ invariant. The only {\it local} gauge symmetry of $\Gamma$, which eventually will have to be fixed, is the non-abelian gauge symmetry: $\Gamma$ enjoys also a {\it global} vector supersymmetry  which, together with the gauge symmetry,  gives rise to the BRST symmetry in (\ref{BRSTfieldsbackgrounds}).

Notice that, with this reinterpretation, the ``ghost field'' associated to the topological supersymmetry is the ghost number +1 combination $\gamma^{(0)}\, A^\ast$, i.e. a {\it composite} field.

$\Gamma$ in (\ref{Actionoffshell}) is invariant under simultaneous transformations of fields and backgrounds.  To obtain the action invariant under {\it rigid} topological supersymmetry we consider the backgrounds which are left invariant under (\ref{BRSTfieldsbackgrounds}) 
\bea
\rmd\,\psi^{(1)}=0 \ ,\qquad  \rmd\, \gamma^{(0)}=0\Leftrightarrow \gamma^{(0)}= \textrm{constant}\equiv \gamma_0 \ .
\eea
One usually restricts oneself to bosonic backgrounds. In this case
\bea
\psi^{(1)}=0\ \ ,
\eea
and the BRST transformations reduce to
\bea
&& s_0\, c = - c^2+ \gamma_0\, \phi \ ,\nonumber\\
&& s_0\, A = - \rmD\,c +\gamma_0 \, A^\ast \ ,\nonumber\\
&& s_0\, \phi = -[c, \phi] \ ,\nonumber\\
&& s_0\, A^\ast= - [c, A^\ast] - \rmD\, \phi \ .
\label{BRSTfieldsrigid}
\eea
By introducing the rescaled fields
\bea
\hat{\phi} \equiv  \gamma_0\, \phi 	 \, \qquad \hat{\psi} \equiv \gamma_0 \,A^\ast \ ,
\eea
with $ \hat{\psi}$ and $\hat{\phi}$ of ghost number 1 and 2 respectively, the BRST transformations (\ref{BRSTfieldsrigid})  become identical to the topological  Yang-Mills BRST transformations (\ref{BRSTtopologicalYM}) and the BRST invariant action coincides with the Witten topological action $\Gamma_{W}$ in (\ref{ActionWitten}).
Notice that, with this approach, the shift on the ghost number of the field $\phi$ does not occur: it is the {\it composite} field $\hat \phi$ that has ghost number $2$. The fundamental field $\phi$ remains of ghost number $0$. This property will be important when we will extend the discussion to more general PSMs.

Let us summarize our logic: we started from  2d YM. To mantain the topological nature   we replaced the 2-dimensional metric and the coupling constant $\epsilon$ with a 2-form background field $f^{(2)}$, at the same time asking that the physics only depends on the cohomology class of $f^{(2)}$.  This entails both extending the BRST gauge transformations to the background (and completing $f^{(2)}$ to a topological $U(1)$ multiplet) and to deform the BRST transformations of the gauge multiplet.
Since the deformed BRST transformations close only up to the equations of motion of the gauge field,  it has been necessary to introduce the auxiliary field $A^\ast$ --- which in the BV formalism would be the antifield of  $A$.  We managed to obtain in this way a BRST invariant theory coupled to topological backgrounds. Theories invariant under rigid supersymmetry are  now obtained by considering the backgrounds which are {\it bosonic} fixed points
of the deformed BRST operator, i.e. $\gamma^{(0)}=\gamma_0$ constant and $\psi^{(1)}=0$.  For $\gamma_0\not=0$ one gets the topological YM  Witten theory and identifies 
the somewhat mysterious topological gaugino $\psi$  of \cite{Witten:1992xu}  as   $\gamma_0\, A^\ast$.

By choosing the point $\gamma_0=0$ in the space of BRST-invariant backgrounds one
recovers the original YM action (\ref{eq:YMactionintro}):  in this limit the topological supersymmetry  collapses
and the BRST symmetry reduces to the pure gauge one, (\ref{eq:sgaugeintro}). By further taking the point $\gamma_0 = f^{(2)} = 0$ one instead obtains back the original BF-model action (\ref{eq:topYMactionintro}). We see therefore that our theory, coupled to the topological backgrounds, provides a {\it topological} extension of the standard 2-dimensional BF-model.

 The fact that  the $\gamma_0=0$ point is  degenerate in the space of backgrounds, gives a conceptual understanding of  why the topological supersymmetry of the standard
YM action is ``hidden''. On generic points $\gamma_0\not=0 $ of the space of backgrounds the topological
supersymmetry is manifest.

\subsection{Superfield formulation }
\label{sub:superfieldYM}

Whatever the point of view  one chooses --- the BV point of view which treats $A^\ast$ as an antifield, or the supersymmetric one in which $A^\ast$ is  an independent field --- it is possible to develop a superfield formulation for the theory coupled to the topological backgrounds. 

To this end, let us introduce the 2-form fields $\phi^\ast$ and $c^\ast$, of ghost number $-1$ and $-2$ respectively. In the BV formalism these fields would be the antifields of the scalars $\phi$ and $c$. We can then introduce the polyforms
\bea
\label{eq:polyformsYMfields}
&& \mathcal{A} \equiv c + A + \phi^\ast \ , \nn \\
&& \Phi \equiv \phi + A^\ast + c^\ast \ ,
\eea
carrying total ghost number (given by the sum of ghost number and form degree) $+1$ and $0$. The background fields are collected in a single polyform
\bea
\label{eq:backgroundsuperfield}
&& \mathbf f \equiv \gamma^{(0)} + \psi^{(1)} + f^{(2)} \ ,
\eea
of total ghost number $+2$; we also introduce the coboundary operator
\bea
\label{eq:coboundarydelta0}
\delta_0 \equiv s_0 + \rmd \, .
\eea
It is now straightforward to see that the relations
\bea
\label{eq:vectorrigidpoly} 
&& \delta_0\, \mathcal{A} + \mathcal{A}^2 = \mathbf{f} \, \Phi \ , \nn \\
&& \delta_0 \, \mathbf{f} = 0 \nn\\
&& \delta_0 \, \Phi+ [ \mathcal{A}, \Phi]=0,
\eea
precisely reproduces the BRST variations (\ref{BRSTfieldsbackgrounds}) together with the BRST transformations for $\phi^\ast$ and $c^\ast$:
\bea
&& s_0 \,  \phi^\ast = - [c ,  \phi^\ast] - F  + \gamma^{(0)}  c^\ast + \psi^{(1)}  A^\ast + f^{(2)} \phi \ , \nn \\
&& s_0 \,  c^\ast = - [c ,  c^\ast] - [ \phi^\ast, \phi] - \rmD  A^\ast \ . 
\eea
Once written in the polyform notation (\ref{eq:vectorrigidpoly}), the geometrical meaning of the coupling to the topological backgrounds is very transparent: when one takes the degenerate point $\mathbf{f} = 0$ the BRST variations for the polyforms $\mathcal{A}$ and $\Phi$ are completely decoupled. On this particular point, one ends up with the standard gauge invariance of the 2-dimensional BF-model. When $\mathbf{f}$ is turned on, the BRST variations for $\mathcal{A}$ and $\Phi$ are coupled: this coupling realizes the topological supersymmetry discussed in Section \ref{sec:topologicalSUSY}.

\newsection{Coupling Poisson sigma models to topological backgrounds}
\label{sec:PSMs}

In this section we will see how the coupling to the topological backgrounds we outlined in Section \ref{sec:2dYM} can be extended to a generic PSM, of which the BF-model is a particular case with a {\it linear} Poisson structure. We will start in Subsection \ref{sec:PSMreview} with a review of the construction of a generic PSM, mainly following \cite{Cattaneo:1999fm}.\footnote{Compared to \cite{Cattaneo:1999fm}, our formulas differ by some signs. This is a consequence of the different conventions we adopt for the (anti)-commutation rules between the BRST operator and the external differential.} Then, in Section \ref{sec:PSMcoupling}, we will introduce the coupling to the topological backgrounds.

\subsection{A review of the Poisson sigma models construction}
\label{sec:PSMreview}

A PSM is a topological sigma model in which the target manifold $M$ is a $d$-dimensional Poisson manifold. As such, $M$ is provided with a Poisson structure, i.e. a bivector $\Pi^{ij} (X) \in \Gamma(\wedge^2 TM)$ satisfying the Jacobi condition
\bea
\label{eq:Poissoncondition}
\left[\Pi \, , \, \Pi \right]_S \equiv \Pi^{il} \partial_l \Pi^{jk} + \Pi^{jl} \partial_l \Pi^{ki} + \Pi^{kl} \partial_l \Pi^{ij} = 0 \ ,
\eea
where $\left[\cdot \, , \, \cdot \right]_S$ denotes the Schouten bracket and $X$ are local coordinates on $M$.

The PSM describes maps from a Riemann-surface $\Sigma$ to the target space $M$. It has two real bosonic (ghost number $0$) fields $X^i$ and $\eta_i$. $X^i$ describes the map from  $\Sigma$ to $M$, i.e. it is represented by $d$ functions $X^i (x)$, where $x$ are collectively the coordinates on $\Sigma$. $\eta_i$ is a 1-form on $\Sigma$ valued in $X^\ast (T^\ast M)$, the pullback of the cotangent bundle on $M$.

The action is given by
\bea
\label{eq:PSMactionzero}
&& \Gamma^0_{\rm PSM} = \int_\Sigma \eta_i \wedge \rmd \, X^i + \frac 12 \, \Pi^{ij}(X) \, \eta_i \wedge \eta_j \ ,
\eea
and, thanks to the Jacobi identity (\ref{eq:Poissoncondition}), it has the gauge invariance
\bea
\label{eq:PSMgaugeinvariancezero}
&& s_0 \, X^i =  - \Pi^{ij} (X) \, \beta_j \ , \nn \\
&& s_0  \, \eta_i = - \rmd \,\beta_i - \partial_i \Pi^{kl} (X) \,\eta_k \beta_l \ , \nn \\
&& s_0 \, \beta_i = - \frac 12 \partial_i \Pi^{jk} (X)\, \beta_j \beta_k \ ,
\eea
where $\beta_i$ is the ghost field (a scalar of ghost number $+1$). Notice that, up to an integration by parts, the action (\ref{eq:PSMactionzero}) and the BRST transformations (\ref{eq:PSMgaugeinvariancezero}), in the special case of a Poisson structure {\it linear} in the coordinates $X^i$, are equivalent to the BF-model action and BRST variations, equations (\ref{eq:topYMactionintro}) and (\ref{eq:sgaugeintro}). This is nothing but the well-known fact that the BF-model is a particular example of PSM.

Contrary to the particular case of the BF-model, the BRST variations (\ref{eq:PSMgaugeinvariancezero}) are {\it not} nilpotent in general. Indeed, using again the Jacobi property (\ref{eq:Poissoncondition}), one finds $s_0^2 \, X^i = s_0^2 \, \beta_i = 0$ and
\bea
\label{eq:PSMnonnilpotencys0}
s_0^2 \, \eta_i = - \frac 12 \partial_i \partial_k \Pi^{rs} (X) \, \beta_r \beta_s \, \bigl( \rmd \, X^k + \Pi^{kj}(X) \, \eta_j \bigr) \ .
\eea
The non-nilpotency of $s_0$ in (\ref{eq:PSMnonnilpotencys0}) is given by a term proportional to the equations of motions. It is therefore necessary to use the BV formalism, or its supersymmetric reinterpretation in terms of auxiliary fields we recalled in Section \ref{sec:topologicalSUSY}. This is the main difference between a generic PSM and the particular case of BF-model: while in the case of the BF-model before the coupling to the topological $U(1)$ multiplet the BRST operator was nilpotent off-shell, for a generic PSM the necessity of the antifields arises from the very beginning.

Let us therefore introduce the antifield $\eta^{\ast i}$. It is a 1-form on $\Sigma$ valued in $T M$ of ghost number $-1$. The BRST transformations get modified to the nilpotent ones
\bea
\label{eq:PSMgaugeinvarianceone}
&& s_0 \, \beta_i = - \frac 12 \partial_i \Pi^{jk} (X)\, \beta_j \beta_k \ , \nn \\
&& s_0  \, \eta_i = - \rmd \, \beta_i - \partial_i \Pi^{kl} (X) \,\eta_k \beta_l - \frac 12 \partial_i \partial_j \Pi^{kl} (X) \, \eta^{\ast j} \beta_k \beta_l \ , \nn \\
&& s_0 \, X^i =  - \Pi^{ij} (X) \, \beta_j \ , \nn \\
&& s_0 \, \eta^{\ast i} = - \rmd \, X^i - \Pi^{ij} (X) \, \eta_j - \partial_k \Pi^{ij}(X) \, \eta^{\ast k} \beta_j \ .
\eea 

Since the BRST transformation rules for $\eta_i$ have been modified by a term including the antifield $\eta^{\ast i}$, the BRST invariance of the action (\ref{eq:PSMgaugeinvariancezero}) is lost
\bea
&& s_0 \, \Gamma^0_{\rm PSM} \neq 0 \ ,
\eea
but it can be restored by adding, as usual, a term quadratic in the antifield $\eta^{\ast i}$
\bea
\label{eq:PSMactionone}
&& \Gamma_{\rm PSM} = \int_{\Sigma} \bigl[ \eta_i \wedge \rmd \, X^i + \frac 12 \, \Pi^{ij}(X) \, \eta_i \wedge \eta_j - \frac 14 \eta^{\ast i} \wedge \eta^{\ast j} \, \partial_i \partial_j \Pi^{kl} (X) \,\beta_k \beta_l \bigr] \ .
\eea
We stress that, both the nilpotency of the BRST operator in (\ref{eq:PSMgaugeinvarianceone}) and the invariance of the action (\ref{eq:PSMactionone}), requires the Jacobi identity (\ref{eq:Poissoncondition}).

A few remarks are now in order. First of all, we notice that the transformations (\ref{eq:PSMgaugeinvarianceone}) do {\it not} have an interpretation as a topological supersymmetry: indeed, even if the antifield $\eta^{\ast i}$ is already present in the BRST variation of the field $\eta_i$, the term
\bea
\label{eq:nonsusytransformations}
- \frac 12 \partial_i \partial_j \Pi^{kl} (X) \, \eta^{\ast j} \beta_k \beta_l \ , 
\eea
is a non-linear term (it is at least cubic). Therefore such a term cannot be interpreted as a topological supersymmetry for the field $\eta_i$. This is not surprising since, as we discussed for the case of the BF-model,  the topological supersymmetry  appears after the coupling to the topological backgrounds.

More importantly we notice that, contrary to the case of the BF-model, by treating the antifield $\eta^{\ast i}$ as an independent field one obtains a theory which is different from the original PSM. Indeed the additional term involving the antifield in (\ref{eq:PSMactionone}) changes the local symmetry content of the theory: since in this term the ghost field $\beta_i$ appears explicitly, we should consider the symmetry carried by the ghost field $\beta_i$ as a {\it global} symmetry, and not as a {\it local} symmetry anymore. We conclude therefore that, if we insist in treating the antifields as independent fields, what we obtain is a theory which is different from the original PSM, in which the antifields are forced to be functionals of the fields during the gauge-fixing and the $\beta_i$ symmetry is local and must be gauge-fixed.\footnote{The author thanks C.~Imbimbo for discussions on this point.}

For this reason, we will discuss how to complete the action  (\ref{eq:PSMactionone}) to the full BV-action which can be eventually gauge-fixed using the standard BV rules. For doing that, let us introduce the antifields $X^{\ast}_i$, a 2-form of ghost number $-1$, and $\beta^{\ast i}$, a 2-form of ghost number $-2$, and let us introduce the superfield formulation.

\subsubsection{Superfield formulation}
\label{sec:PSMsuperfield}

Let us  define the polyforms
\bea
\label{eq:PSMpolyform}
&& \widetilde{  X}^i \equiv X^i + \eta^{\ast i} + \beta^{\ast i} \ , \nn \\
&& \widetilde{  \eta}_i \equiv \beta_i + \eta_i + X^{\ast}_i \ , 
\eea
of total ghost number $0$ and $+1$, respectively. Notice that $\widetilde{X}^i$ is a polyform on $\Sigma$ valued on $TM$ and $\widetilde{ \eta}_i$ is a polyform on $\Sigma$ valued on $T^\ast M$. Moreover, the Poisson bivector $\Pi^{ij} (X)$ is promoted to the polyform of total ghost number $0$
\bea
\label{eq:Poissonsuperfield}
&& \Pi^{ij} (\widetilde{  X}) \equiv \bigr( \Pi^{ij} (X) \bigr) + \bigl( \partial_k \Pi^{ij} (X) \, \eta^{\ast k} \bigr) + \bigl( \frac 12 \partial_k \partial_l \Pi^{ij}(X) \, \eta^{\ast k} \wedge \eta^{\ast l} + \partial_k \Pi^{ij} (X) \, \beta^{\ast k} \bigr) \ ,
\eea
which satisfies the elegant relation
\bea
\label{eq:jacobipolyform}
&& \Pi^{ i l} (\widetilde X) \partial_l \Pi^{jk} (\widetilde X) + \Pi^{ j l} (\widetilde X) \partial_l \Pi^{ki} (\widetilde X) + \Pi^{ k l} (\widetilde X) \partial_l \Pi^{ij} (\widetilde X) = 0 \ ,
\eea
formally identical to the Jacobi condition (\ref{eq:Poissoncondition}).

The BRST variations of both the fields and antifields can be described in a compact notation as
\bea
\label{eq:PSMsuperfieldBRST}
&& \delta_0 \, \widetilde{  X}^i = - \Pi^{ij} (\widetilde{  X}) \, \widetilde{  \eta}_i \ , \nn \\
&& \delta_0 \, \widetilde{  \eta}_i = - \frac 12 \partial_i \Pi^{jk} (\widetilde{  X}) \, \widetilde{  \eta}_j \widetilde{  \eta}_k \  ,
\eea
where, as in (\ref{eq:coboundarydelta0}), $\delta_0$ is given by $\delta_0 \equiv s_0 + \rmd$. Written in components (\ref{eq:PSMsuperfieldBRST}) read
\bea
\label{eq:PSMgaugeinvariancefinal}
&& s_0 \, \beta_i = - \frac 12 \partial_i \Pi^{jk} (X)\, \beta_j \beta_k \ , \nn \\
&& s_0  \, \eta_i = - \rmd \, \beta_i - \partial_i \Pi^{kl} (X) \,\eta_k \beta_l - \frac 12 \partial_i \partial_j \Pi^{kl} (X) \, \eta^{\ast j} \beta_k \beta_l \ , \nn \\
&& s_0 \, X^{\ast}_i = - \rmd \, \eta_i - \partial_i \Pi^{kl}(X) \, X^{\ast}_k \beta_l - \partial_i\partial_j \Pi^{kl}(X)\, \eta^{\ast j} \wedge \eta_k \beta_l - \frac 12 \partial_i\Pi^{kl}(X)\, \eta_k \wedge \eta_l + \nn \\
&& \qquad  \; \; \; \; \; - \frac 14 \partial_i \partial_j \partial_p \Pi^{kl} (X) \, \eta^{\ast j} \wedge \eta^{\ast p} \beta_k \beta_l - \frac 12 \partial_i \partial_j \Pi^{kl} (X) \, \beta^{\ast j} \beta_k \beta_l  \ , \nn \\
&& s_0 \, X^i =  - \Pi^{ij} (X) \, \beta_j \ , \nn \\
&& s_0 \, \eta^{\ast i} = - \rmd \, X^i - \Pi^{ij} (X) \, \eta_j - \partial_k \Pi^{ij}(X) \, \eta^{\ast k} \beta_j \ , \nn \\
&& s_0 \, \beta^{\ast i} = - \rmd \, \eta^{\ast i} - \Pi^{ij}(X)\, X^{\ast}_j - \frac 12 \partial_k \partial_l \Pi^{ij}(X)\, \eta^{\ast k} \wedge \eta^{\ast l} \beta_j  + \nn \\
&& \qquad  \; \; \; \; \, - \partial_k \Pi^{ij} (X) \, \eta^{\ast k} \wedge \eta_j - \partial_k \Pi^{ij} (X) \, \beta^{\ast k} \beta_ j \ .
\eea

\subsubsection{The BV action}
\label{sec:BVaction}

As anticipated, let us complete the action (\ref{eq:PSMactionone}) by constructing the full BV action for the PSM. To this end, we have to consider the canonical piece, that we wrote schematically in (\ref{eq:canonicalscheme}) and that in this case reads
\bea
\label{eq:canonicalPoisson}
&& \Gamma_{\rm can} = - \int_{\Sigma} \bigl(  \beta^{\ast i} s_0 \, \beta_i  + \eta^{\ast i} \wedge (s_0 \, \eta_i) + X^{\ast}_i s_0 \, X^i\bigr) \ .
\eea
By a simple computation we observe that $ \Gamma_{\rm can}$ is BRST invariant:
\bea
&& s_0 \, \Gamma_{\rm can} = 0 \ ,
\eea
and so we conclude that, the full BV action\footnote{From which, using the usual BV rules,  we obtain the BRST variations of the fields and antifields
\bea
s_0 \, \Phi = - \frac{\overrightarrow{\partial} \, \Gamma_{\rm BV}}{\partial \Phi^\ast} \ , \qquad s_0 \, \Phi^\ast = - \frac{\overrightarrow{\partial} \, \Gamma_{\rm BV}}{\partial \Phi} \ . 
\eea}
\bea
\Gamma_{\rm BV} = \Gamma_{\rm PSM} + \Gamma_{\rm can} \ ,
\eea
is given by the sum of two pieces which are  {\it separately} BRST invariant
\bea
\label{eq:BRSTinvariancefullBV}
&& s_0 \, \Gamma_{\rm BV} = s_0 \, \Gamma_{\rm PSM} = s_0 \, \Gamma_{\rm can} = 0 \ .
\eea
Thanks to the property (\ref{eq:BRSTinvariancefullBV}), when in the following we will discuss the invariance of the action we will sistematically restrict our attention to the non-canonical term only, since the canonical term will be automatically BRST invariant.

\subsection{The coupling to topological backgrounds}
\label{sec:PSMcoupling}

Having described the general theory of PSMs, it becomes natural to couple the PSM to the topological backgrounds (\ref{eq:backgroundsuperfield}), similarly to what we did for the BF-model (\ref{eq:topYMactionintro}). As we have seen, this coupling provides a topological reformulation of 2-dimensional YM and makes manifest the topological supersymmetry. 

Given the BRST transformations for 2-dimensional YM written in the polyform notation (\ref{eq:vectorrigidpoly}), and given the similarities with the transformations (\ref{eq:PSMsuperfieldBRST}) for a generic PSM, one would guess that the second equation in (\ref{eq:PSMsuperfieldBRST}) should be modified by a term involving $\mathbf f$ and the superfield $\widetilde{  X}^i$. However, this cannot be the correct way to perform the coupling: indeed $\widetilde{  X}^i$ and $\widetilde{\eta}_i$ are valued in $TM$ and $T^\ast M$ respectively, and therefore they cannot be related directly in the BRST variations. 

On the other hand, the deformation term appearing in (\ref{eq:nontoptermintro}) is constructed via a {\it Casimir function}: given a certain Poisson manifold $M$, with the corresponding Poisson bivector $\Pi^{ij}(X)$, a function $f(X) \in  C^{\infty}(M)$ is said to be Casimir if
\bea
\label{eq:Casimirdefinition}
&& \bigl[\Pi \, , \, f \bigr]_S \equiv \Pi^{ij}(X)\, \partial_j f(X) = 0 \ .
\eea
The term $\Tr \, (\phi^2)$ is an example of Casimir function for the particular case of the BF-model.

Let us take a Casimir function $C(X)$ on $M$. Thanks to the Casimir property (\ref{eq:Casimirdefinition}) it satisfies
\bea
\label{eq:BRSTcasimir}
&& s_0 \, C(X) = \partial_i C(X) \, s_0\, X^i = - \partial_i C(X) \, (\Pi^{ij}(X) \, \beta_j) = 0 \ ,
\eea
i.e. it is BRST invariant. Let us also modify the PSM action (\ref{eq:PSMactionone}) to\footnote{The action analogous to (\ref{eq:YMactionintro}) for a generic PSM, i.e. the action in which the Casimir function is added in a way which breaks the topological invariance, has been already considered in the past \cite{Schwarzweller:2001pw}. We stress that our approach, instead, preserves the topological nature of the model by considering the background multiplet $\mathbf f$. }
\bea
\label{eq:PSMcoupledone}
&& \Gamma^C_{\rm PSM} [f^{(2)}, C (X)] = \int_{\Sigma} \bigl[ \eta_i \wedge \rmd \, X^i + \frac 12 \, \Pi^{ij}(X) \, \eta_i \wedge \eta_j + \nn \\
&& \qquad \qquad - \frac 14 \eta^{\ast i} \wedge \eta^{\ast j} \, \partial_i \partial_j \Pi^{kl} (X) \,\beta_k \beta_l  - f^{(2)} \, C(X)\bigr] \ .
\eea
Again, the BRST invariance is lost
\bea
&& s_0 \, \Gamma^C_{\rm PSM} [f^{(2)}, C (X)] = + \int_\Sigma \rmd \, \psi^{(1)} C(X) = - \int_\Sigma \partial_i C(X) \, \psi^{(1)} \wedge \rmd \, X^i \ ,
\eea
but it can be restored by modifying the BRST variation of $\eta_i$
\bea
\label{eq:BRSTPSMcoupledeta}
s_0 \, \eta_i = - \rmd \, \beta_i - \partial_i \Pi^{kl} (X) \,\eta_k \beta_l - \frac 12 \partial_i \partial_j \Pi^{kl} (X) \, \eta^{\ast j} \beta_k \beta_l + \partial_i C(X) \, \psi^{(1)} \ .
\eea
With this modification the action is BRST invariant
\bea
&& s_0 \, \Gamma^C_{\rm PSM} [f^{(2)}, C (X)] = 0 \ ,
\eea
but $s_0$ is not nilpotent anymore on the field $\eta_i$
\bea
&& s^2_0\, \eta_i = + \partial_i \Pi^{kl}(X)\, \partial_k C(X)\, \psi^{(1)} \beta_l +  \Pi^{kl}(X)\, \partial_k \partial_i C(X)\, \psi^{(1)} \beta_l - \partial_i C(x)\, \rmd \, \gamma^{(0)} = \nn \\
&& \qquad \, =  - \partial_i C (X) \, \rmd \, \gamma^{(0)} \ ,
\eea
where, again, we made use of the Casimir property (\ref{eq:Casimirdefinition})
\bea
&& \partial_i \Pi^{kl}(X)\, \partial_k C(X)\, \psi^{(1)} \beta_l +  \Pi^{kl}(X)\, \partial_k \partial_i C(X) = \partial_i (\Pi^{kl} (X) \partial_k C(X)) \beta_l = 0 .
\eea

Similarly to what we did for the 2-dimensional BF-model, let us introduce a deformation term in the BRST variation of $\beta_i$
\bea
&& s_0\, \beta_i = - \frac 12 \partial_i \Pi^{kl}(X) \, \beta_k \beta_l + \gamma^{(0)} \partial_i C(X) \ .
\eea
The nilpotency of $s_0$ on $\beta_i$ and $X^i$ is preserved
\bea
&& s^2_0 \, \beta_i = \gamma^{(0)} \partial_i \bigl(\Pi^{kl}(X) \, \partial_l C(X) \bigr) \beta_k = 0 \ , \nn \\
&& s^2_0 \, X^i = - \gamma^{(0)} \Pi^{ij}(X)\, \partial_j C(X) = 0 \ ,
\eea
whereas, exactly as we did for the 2-dimensional BF-model, the nilpotency of the BRST operator on $\eta_i$ requires a new term {\it linear} in the antifield $\eta^{\ast j}$:
\bea
&& s_0 \, \eta_i  = - \rmd \, \beta_i - \partial_i \Pi^{kl} (X) \,\eta_k \beta_l - \frac 12 \partial_i \partial_j \Pi^{kl} (X) \, \eta^{\ast j} \beta_k \beta_l + \nn \\
&& \qquad \quad + \partial_i C(X) \, \psi^{(1)} + \gamma^{(0)} \partial_i \partial_j C(X) \, \eta^{\ast j} \ .
\eea
Summarising, the nilpotent BRST transformations on the fields $\beta_i$, $\eta_i$, $X^i$ and $\eta^{\ast i}$ coupled to topological backgrounds are
\bea
\label{eq:BRSTPSMcoupled}
&& s_0 \, \beta_i = - \frac 12 \partial_i \Pi^{jk} (X)\, \beta_j \beta_k + \gamma^{(0)} \partial_i C(X) \ , \nn \\
&& s_0  \, \eta_i = - \rmd \, \beta_i - \partial_i \Pi^{kl} (X) \,\eta_k \beta_l - \frac 12 \partial_i \partial_j \Pi^{kl} (X) \, \eta^{\ast j} \beta_k \beta_l + \nn \\
&& \qquad \quad + \partial_i C(X) \, \psi^{(1)} + \gamma^{(0)} \partial_i \partial_j C(X) \, \eta^{\ast j} \ , \nn \\
&& s_0 \, X^i =  - \Pi^{ij} (X) \, \beta_j \ , \nn \\
&& s_0 \, \eta^{\ast i} = - \rmd \, X^i - \Pi^{ij} (X) \, \eta_j - \partial_k \Pi^{ij}(X) \, \eta^{\ast k} \beta_j \ , \nn \\
&& s_0\, f^{(2)} = - \rmd \, \psi^{(1)} \ ,\nonumber\\
&& s_0\, \psi^{(1)} = - \rmd \, \gamma^{(0)} \ ,\nonumber\\
&& s_0\, \gamma^{(0)}=0 \ .
\eea
Notice that, in the particular case of a linear Poisson structure and a quadratic Casimir function $C(X)$, we obtain the same expressions we got for 2-dimensional YM (\ref{BRSTfieldsbackgrounds}).

Since we modified the BRST variations of the field $\eta_i$ by a term involving the antifield $\eta^{\ast i}$, the action (\ref{eq:PSMcoupledone}) is not BRST invariant anymore under BRST transformations of both the fields and backgrounds, and we must add another term {\it quadratic} in the antifields
\bea
\label{eq:PSMcoupledtwo}
&& \Gamma^C_{\rm PSM} [f^{(2)}, \gamma^{(0)} ,  C (X)] = \int_{\Sigma} \bigl[ \eta_i \wedge \rmd \, X^i + \frac 12 \, \Pi^{ij}(X) \, \eta_i \wedge \eta_j + \nn \\
&& \qquad  - \frac 14 \eta^{\ast i} \wedge \eta^{\ast j} \, \partial_i \partial_j \Pi^{kl} (X) \,\beta_k \beta_l  - f^{(2)} \, C(X) + \frac 12 \gamma^{(0)} \, \eta^{\ast i} \wedge \eta^{\ast j} \, \partial_i \partial_j C(X) \bigr] \ .
\eea

What we obtained is the equivalent, for a generic PSM, of the discussion explained in Section \ref{sec:topologicalSUSY}: we have taken the topological PSM and we have coupled it to the topological backgrounds {\it and} to the Casimir function $C(X)$. In doing this extension we have obtained the new action and BRST transformations, formulas (\ref{eq:PSMcoupledtwo}) and (\ref{eq:BRSTPSMcoupled}). We have seen that these modifications, in the case of a linear Poisson structure and quadratic Casimir, give the transition from the BF-model to a reformulation of 2-dimensional YM and they provide the topological supersymmetry we discussed in Section \ref{sec:topologicalSUSY}. In the case at hand, we see that the BRST variation of $\eta_i$ acquires a term
\bea
\label{eq:susytopologicalPSM}
\gamma^{(0)} \partial_i \partial_j C(X) \, \eta^{\ast j} \ ,
\eea
which, for Casimir functions $C(X)$ which are {\it quadratic} in the local coordinates $X^i$ (or, at least, that can be expanded in series and contain a quadratic piece in the expansion), can be interpreted as the topological supersymmetry we were looking for. Therefore we see again that, also for generic PSMs (beyond the case of the BF-model), the coupling to the topological $U(1)$ multiplet introduces a topological supersymmetry in the model. However it is interesting to observe that, contrary to the case of the BF-model, in a generic PSM the $\beta_i$-symmetry and the topological supersymmetry controlled by the term (\ref{eq:susytopologicalPSM}) are mixed together.

Let us complete the description of the BRST transformations by introducing, as usual, the polyform notation and considering also the antifields $X^{\ast}_i$ and $\beta^{\ast i}$.

\subsubsection{Superfield formulation for the coupled theory}
\label{sec:PSMcoupledsuperfield}

Beyond the polyforms for the dynamical fields (\ref{eq:PSMpolyform}), for the topological backgrounds (\ref{eq:backgroundsuperfield}) and for the Poisson bivector (\ref{eq:Poissonsuperfield}) we also introduce a polyform, of total ghost number $0$, corresponding to the Casimir function $C$:
\bea
\label{eq:Casimirpolyform}
&& C (\widetilde X) \equiv C(X) + \bigl(\partial_i C(X)\, \eta^{\ast i}\bigr) + \bigl(\frac 12 \partial_i \partial_j C(X)\,\eta^{\ast i} \wedge \eta^{\ast j} + \partial_i C(X) \, \beta^{\ast i} \bigr)  \ .
\eea

The BRST variations for the fields and the backgrounds are again rewritten in terms of the coboundary operator $\delta_0 $ of (\ref{eq:coboundarydelta0}). They reads
\bea
\label{eq:BRSTvariationsPSMcoupledpolyforms}
&& \delta_0 \, \mathbf{f} = 0 \ , \nn \\
&& \delta_0 \, \widetilde X^i = - \Pi^{ij}(\widetilde X) \, \widetilde \eta_j \ , \nn \\
&& \delta_0 \, \widetilde \eta_i = - \frac 12 \, \partial_i \Pi^{jk} (\widetilde X)\, \widetilde \eta_j \widetilde \eta_k + \mathbf{f} \, \partial_i C(\widetilde X) \ .
\eea
Notice that, in the polyform notation,  the Casimir condition (\ref{eq:Casimirdefinition}) is rewritten in terms of the superfields $\Pi^{ij}(\widetilde X)$ and $C(\widetilde X)$ in the elegant form
\bea
\label{eq:PSMconstraintspolyforms}
&& \Pi^{ij} (\widetilde X) \partial_j C(\widetilde X) = 0 \ ,
\eea
formally identical to (\ref{eq:Casimirdefinition}).

Again, when expressed in the polyform notation, the meaning of the deformation is much more transparent: via the topological background $\mathbf f$, the two superfields $\widetilde X^i$ and $\widetilde \eta^i$ get coupled. This coupling realizes, at least for Casimir functions $C(X)$ that can be expanded in series (and that have a quadratic term in the expansion), the topological supersymmetry of the deformed model. 

Written in components, the BRST variations for the fields and the backgrounds reproduces the transformations (\ref{eq:BRSTPSMcoupled}), completed with the BRST variations of the fields $X^{\ast }_i$ and $\beta^{\ast i}$
\bea
\label{eq:BRSTPSMcoupledfinal}
&& s_0 \, \beta_i = - \frac 12 \partial_i \Pi^{jk} (X)\, \beta_j \beta_k + \gamma^{(0)} \partial_i C(X) \ , \nn \\
&& s_0  \, \eta_i = - \rmd \beta_i - \partial_i \Pi^{kl} (X) \,\eta_k \beta_l - \frac 12 \partial_i \partial_j \Pi^{kl} (X) \, \eta^{\ast j} \beta_k \beta_l + \nn \\
&& \qquad \quad + \partial_i C(X) \, \psi^{(1)} + \gamma^{(0)} \partial_i \partial_j C(X) \, \eta^{\ast j} \ , \nn \\
&& s_0 \, X^{\ast}_i = - \rmd \eta_i - \partial_i \Pi^{kl}(X) \, X^{\ast}_k \beta_l - \partial_i\partial_j \Pi^{kl}(X)\, \eta^{\ast j} \wedge \eta_k \beta_l - \frac 12 \partial_i\Pi^{kl}(X)\, \eta_k \wedge \eta_l + \nn \\
&& \qquad  \; \; \; \; \; - \frac 14 \partial_i \partial_j \partial_p \Pi^{kl} (X) \, \eta^{\ast j} \wedge \eta^{\ast p} \beta_k \beta_l - \frac 12 \partial_i \partial_j \Pi^{kl} (X) \, \beta^{\ast j} \beta_k \beta_l  + \nn \\
&& \qquad  \; \; \; \; \; + f^{(2)} \, \partial_i C(X) + \partial_i \partial_k C(X) \, \bigl( \gamma^{(0)} \beta^{\ast k} + \eta^{\ast k} \wedge \psi^{(1)} \bigr) + \frac 12 \, \gamma^{(0)} \partial_i \partial_k \partial_j C(X) \, \eta^{\ast k} \wedge \eta^{\ast j } \ , \nn \\
&& s_0 \, X^i =  - \Pi^{ij} (X) \, \beta_j \ , \nn \\
&& s_0 \, \eta^{\ast i} = - \rmd X^i - \Pi^{ij} (X) \, \eta_j - \partial_k \Pi^{ij}(X) \, \eta^{\ast k} \beta_j \ , \nn \\
&& s_0 \, \beta^{\ast i} = - \rmd \eta^{\ast i} - \Pi^{ij}(X)\, X^{\ast}_j - \frac 12 \partial_k \partial_l \Pi^{ij}(X)\, \eta^{\ast k} \wedge \eta^{\ast l} \beta_j  + \nn \\
&& \qquad  \; \; \; \; \, - \partial_k \Pi^{ij} (X) \, \eta^{\ast k} \wedge \eta_j - \partial_k \Pi^{ij} (X) \, \beta^{\ast k} \beta_ j \ , \nn \\
&& s_0\, f^{(2)} = - \rmd\, \psi^{(1)} \ ,\nonumber\\
&& s_0\, \psi^{(1)} = - \rmd\, \gamma^{(0)} \ ,\nonumber\\
&& s_0\, \gamma^{(0)}=0 \ .
\eea

\subsection{The polyform for the action}
\label{sec:actionPSMcoupledpolyform}

We have seen that the action for the PSM coupled to the topological backgrounds is given by the formula (\ref{eq:PSMcoupledtwo}). It also satisfies the so-called {\it descent equation}. Let us consider a generic observable of the theory, given by a 2-form of ghost number $p$, $\mathcal{O}^{(2)}_p$. It is possible to complete $\mathcal{O}^{(2)}_p$ to a polyform, of total ghost number $p+2$, that we call $\mathbf{\mathcal{O}}^{p+2}$
\bea
\mathbf{\mathcal{O}}^{p+2} \equiv \mathcal{O}^{(2)}_p + \mathcal{O}^{(1)}_{p + 1} + \mathcal{O}^{(0)}_{p + 2} \ ,
\eea
where $\mathcal{O}^{(1)}_{p + 1}$ and $\mathcal{O}^{(0)}_{p + 2}$, are taken to be solutions of the system of equations (the descent equation)
\bea
\label{eq:descentequations}
&& s_0\, \mathcal{O}^{(2)}_p = - \rmd \, \mathcal{O}^{(1)}_{p + 1} \ , \nn \\
&& s_0\, \mathcal{O}^{(1)}_{p+1} = - \rmd \, \mathcal{O}^{(0)}_{p + 2} \ , \nn \\
&& s_0 \, \mathcal{O}^{(0)}_{p + 2} = 0 \ ,
\eea
that can be compactly rewritten, in terms of the polyform $\mathbf{\mathcal{O}}^{p+2}$ and of the coboundary operator $\delta_0$  as
\bea
\label{eq:descentequationcompact}
&& \delta_0 \, \mathbf{\mathcal{O}}^{p+2} = 0 \ .
\eea

Let us take $\mathcal{O}^{(2)}_0$ the ghost number $0$ observable given by the topological action
\bea
\label{eq:twoformactioncoupled}
&& \mathcal{O}^{(2)}_0 \equiv \eta_i \wedge \rmd \, X^i + \frac 12 \, \Pi^{ij}(X) \, \eta_i \wedge \eta_j + \nn \\
&& \qquad  - \frac 14 \eta^{\ast i} \wedge \eta^{\ast j} \, \partial_i \partial_j \Pi^{kl} (X) \,\beta_k \beta_l  - f^{(2)} \, C(X) + \frac 12 \gamma^{(0)} \, \eta^{\ast i} \wedge \eta^{\ast j} \, \partial_i \partial_j C(X) \ .
\eea
By computing the BRST variation of $\mathcal{O}^{(2)}_0$, one observes that it gives rise to the polyform, of total ghost number $2$,
\bea
\mathbf{\mathcal{O}}^{2} \equiv \mathcal{O}^{(2)}_0 + \mathcal{O}^{(1)}_{ 1} + \mathcal{O}^{(0)}_{2} \ ,
\eea
where the descendants solving the descent equations (\ref{eq:descentequations}) are
\bea
\label{eq:descentPSMcoupledaction}
&& \mathcal{O}^{(1)}_1 = \beta_i \, \rmd \, X^i - \psi^{(1)} C (X) \ , \nn \\
&& \mathcal{O}^{(0)}_2 =  - \frac 12 \Pi^{ij} (X)\, \beta_i \beta_j - \gamma^{(0)} C(X) \ .
\eea
However, the 0-form $\mathcal{O}^{(0)}_2$ can be also completed to the {\it algebraic} polyform
\bea
\label{eq:polyformactioncoupledalgebraic}
\widetilde{\mathcal{O}}^2 \equiv  - \frac 12 \, \Pi^{ij}(\widetilde X) \widetilde \eta_i \widetilde \eta_j - \mathbf f \, C(\widetilde X) \ ,
\eea
which also solves the descent equation (\ref{eq:descentequationcompact}), and whose 1-form and 2-form parts are given by 
\bea
\label{eq:tildepolyformcoupled}
&& \widetilde{\mathcal{O}}^{(1)}_1 = - \frac 12 \partial_k \Pi^{ij}(X) \, \eta^{\ast k} \beta_i \beta_j - \Pi^{ij}(X) \, \beta_i \eta_j - \psi^{(1)} C(X) - \gamma^{(0)} \partial_i C(X) \, \eta^{\ast i} \ , \nn \\
&& \widetilde{\mathcal{O}}^{(2)}_0 = - \frac 14 \, \partial_l \partial_k \Pi^{ij} (X) \, \eta^{\ast l} \wedge \eta^{\ast k} \beta_i \beta_j - \frac 12 \, \partial_k \Pi^{ij}(X) \, \beta^{\ast k} \beta_i \beta_j - \partial_k \Pi^{ij}(X) \, \beta_i \eta_j \wedge \eta^{\ast k} + \nn \\
&& - \frac 12 \, \Pi^{ij}(X) \, \eta_i \wedge \eta_j - \Pi^{ij}(X)\, \beta_i X^{\ast }_j - f^{(2)} C(X) - \partial_i C(X) \psi^{(1)} \wedge \eta^{\ast i} + \nn \\
&& - \frac 12 \, \gamma^{(0)} \partial_k \partial_i C(X) \, \eta^{\ast k} \wedge \eta^{\ast i} - \gamma^{(0)} \partial_i C(X) \, \beta^{\ast i} \ .
\eea

Summarising, we have found that the 0-form observable 
\bea
&& \mathcal{O}^{(0)}_2 =  - \frac 12 \Pi^{ij} (X)\, \beta_i \beta_j - \gamma^{(0)} C(X) \ ,
\eea
can be completed to two different polyforms, both solving the descent equation (\ref{eq:descentequationcompact}). The first one, that we called $\mathcal{O}^2$, is given by (\ref{eq:twoformactioncoupled}) and (\ref{eq:descentPSMcoupledaction}); in particular it includes, at the 2-form level, the topological action (\ref{eq:PSMcoupledtwo}). The second one, that we called $\widetilde{\mathcal{O}}^2$, is totally algebraic and it is given by (\ref{eq:polyformactioncoupledalgebraic}) and (\ref{eq:tildepolyformcoupled}). The two polyforms are related by a $\delta_0$-trivial cocycle 
\bea
\label{eq:relationPSMcoupledpolyform}
&& \mathcal{O}^2 = \widetilde{\mathcal{O}}^2 + \delta_0 (\alpha^\prime) \ , \nn \\
&& \alpha^\prime = \eta_i \wedge \eta^{\ast i} + \beta_i \beta^{\ast i} +\beta_i \eta^{\ast i} \ ,
\eea
where $\alpha^\prime$ has total ghost number $+1$. We conclude therefore that $\widetilde{\mathcal{O}}^2$ and $\mathcal{O}^2$ are BRST equivalent.

\subsection{The observables}
\label{sec:Observablescoupled}

Let us discuss the observables of the coupled theory (\ref{eq:BRSTPSMcoupledfinal}). Without introducing the deformation controlled by  $C(X)$, a class of observables has been considered in \cite{Bonechi:2007ar} and they go in correspondence with the elements of the {\it Poisson cohomology} of the target space $M$. We will review the analysis of \cite{Bonechi:2007ar} and we will see how it gets modified in the deformed model.

To this end, let us review the concept of Poisson cohomology. It is well-known that, for a Poisson manifold $(M \, , \, \Pi)$, the Schouten bracket $\left[\cdot \ , \ \cdot \right]_S$ satisfies the relation (see, for example, \cite{dufourzungbook})
\bea
\label{eq:schoutencohomology}
\left[ \Pi \ , \left[ \Pi , A\right]_S \right]_S = 0 \ ,
\eea
where $A$ is an arbitrary multivector field, i.e. a section of $\wedge^p TM$ with $p$ generic. (\ref{eq:schoutencohomology}) states that the operator
\bea
\left[ \Pi \ , \ \cdot \right]_S \ ,
\eea
which sends $p$-vector fields to $p+1$-vector fields, is nilpotent. Therefore it makes sense to define the cohomology groups (called the {\it Poisson cohomology groups}) 
\bea
H^p_{\Pi} (M) = \frac{\mathrm{ker} \bigl( \left[ \Pi \ , \ \cdot \right]_S \ : \ T^{p}(M) \rightarrow  T^{p+1}(M) \bigr) }{\mathrm{Im} \bigl( \left[ \Pi \ , \ \cdot \right]_S \ : \ T^{p -1}(M) \rightarrow  T^{p}(M) \bigr)} \ .
\eea
Recall that, given a Poisson manifold $(M \, , \, \Pi)$, there is a natural homomorphism between the de Rham cohomology and the Poisson cohomology. However, except for the very special case in which $M$ is symplectic, this homomorphism is not an isomorphism and actually the Poisson cohomology groups can be very big, even infinite-dimensional.

Let us take $w(X) \in \Gamma (\wedge^p TM)$ and construct the superfield, of total ghost number $p$
\bea
\label{eq:observablesrigidpolyform}
\mathcal{O}^p = \mathcal{O}^{(2)}_{p-2} + \mathcal{O}^{(1)}_{p-1} + \mathcal{O}^{(0)}_{p} \equiv w^{i_1 \cdots i_p} (\widetilde X) \widetilde{\eta}_{i_1} \cdots \widetilde{\eta}_{i_p} \ ,
\eea
whose components read
\bea
\label{eq:observablesrigidcomponents}
&& \mathcal{O}^{(0)}_p = w^{i_1 \cdots i_p} (X)\,\beta_{i_1} \cdots \beta_{i_p} \ , \nn \\
&& \mathcal{O}^{(1)}_{p-1} = \partial_k  w^{i_1 \cdots i_p} (X)\, \eta^{\ast k}\beta_{i_1} \cdots \beta_{i_p} + p\, w^{i_1 i_2 \cdots i_p} (X)\, \eta_{i_2}\beta_{i_2} \cdots \beta_{i_p} \ , \nn \\
&& \mathcal{O}^{(2)}_{p-2} = \frac 12 \partial_l \partial_k  w^{i_1 \cdots i_p} (X)\, \eta^{\ast l} \wedge \eta^{\ast k}\beta_{i_1} \cdots \beta_{i_p} + \partial_k  w^{i_1 \cdots i_p} (X)\, \beta^{\ast k}\beta_{i_1} \cdots \beta_{i_p} + \nn \\
&& \qquad \ \, + p\, \partial_k  w^{i_1 i_2 \cdots i_p} (X)\, \eta^{\ast k} \wedge \eta_{i_1} \beta_{i_2} \cdots \beta_{i_p} + p \, w^{i_1 i_2 \cdots i_p} (X)\, X^\ast_{i_1} \beta_{i_2} \cdots \beta_{i_p} + \nn \\
&& \qquad \ \, + \frac{p (p-1)}{2} \, w^{i_1 i_2 i_3 \cdots i_p} (X) \,\eta_{i_1} \wedge \eta_{i_2} \beta_{i_3} \cdots \beta_{i_3} \ ,
\eea
and that generalizes the expressions (\ref{eq:tildepolyformcoupled}) to ghost numbers different from $2$.

Let us compute the BRST variation of (\ref{eq:observablesrigidpolyform}). Using the BRST variations for the superfields, formula (\ref{eq:BRSTvariationsPSMcoupledpolyforms}), we get
\bea
\label{eq:BRSTvariationobservablesrigidpolyform}
&& s_0 \, \mathcal{O}^p \equiv s_0\, (w^{i_1 \cdots i_p} (\widetilde X) \widetilde{\eta}_{i_1} \cdots \widetilde{\eta}_{i_p}) =  \nn \\
&& = - \rmd \, (w^{i_1 \cdots i_p} (\widetilde X) \widetilde{\eta}_{i_1} \cdots \widetilde{\eta}_{i_p}) - \frac 12 \left(\left[\Pi (\widetilde X) \ , \ w (\widetilde X) \right]_S \right)^{j i_1 \cdots i_p} \widetilde \eta_j \widetilde \eta_{i_1} \cdots \widetilde \eta_{i_p} + \nn \\
&& + \mathbf f \left( \left[ w (\widetilde X) \ , \ C (\widetilde X)\right]_S \right)^{i_2 \cdots i_p} \widetilde \eta_{i_2} \cdots \widetilde \eta_{i_p} \ .
\eea
The second line in (\ref{eq:BRSTvariationobservablesrigidpolyform}) is the BRST variation one obtains in the undeformed model: from this expression we see that, in the underformed model, the observables are in correspondence with the elements of the Poisson cohomology of $M$; in other words $\mathcal{O}^p$ defines an observable for the underformed model iff the multivector field $w(X) \in \Gamma (\wedge^p TM)$ lies in the Poisson cohomology group $H^p_{\Pi} (M)$. On the other hand, the second line in (\ref{eq:BRSTvariationobservablesrigidpolyform}) tells us that, in the deformed model, the fact that $w(X)$ lies in the Poisson cohomology is a necessary condition for $\mathcal{O}^p$ being an observable but it is not sufficient: one has also to require that the Schouten bracket 
\bea
\label{eq:schoutenconstraint}
\left[ w ( X) \ , \ C ( X)\right]_S \ , 
\eea 
vanishes. 

In other words, we conclude that the observables of the deformed theory are in correspondence with the elements of the Poisson cohomology of $M$ that also commute with the Casimir function $C(X)$. Notice that in the particular case of $M$ being a symplectic manifold the additional requirement (\ref{eq:schoutenconstraint}) is automatically satisfied, since the Casimir functions $C(X)$ are simply constant in the local coordinates.

To conclude this subsection we note that an observable $\mathcal{O}^p$ of total ghost number $p$ can be transformed into a {\it composite} observable, of total ghost number $p+2 n$, obtained by dressing $\mathcal{O}_p$ with powers of the background $\mathbf f$:
\bea
\label{eq:dressedobservables}
\mathcal{O}^{p+2 n}_{\mathbf f^n} \equiv (\mathbf f)^n \, \mathcal{O}^{p} \ .
\eea
A simple example of this kind of observables is given by 
\bea
\mathbf f \, C(\widetilde X) \ ,
\eea
which appears in the topological action (\ref{eq:polyformactioncoupledalgebraic}).
\section{The coupling to topological gravity}
\label{sec:couplinggravity}

In this Section we will explain how the PSMs (deformed or not) can be further coupled to 2-dimensional topological gravity.\footnote{The coupling to topological gravity for the special case of a BF-model (and its deformation) has been already discussed in \cite{Bae:2015eoa}, here we will see that exactly the same construction can be exported to generic PSMs.} This is the first step to study topological strings in propagation on a Poisson manifold $M$, a problem that we hope to address in future works. Another motivation to study the coupling to topological gravity is the following: as already remarked, it has been observed in \cite{Imbimbo:2014pla} and \cite{Bae:2015eoa} that the topological theory coupled to rigid topological gravity (and, in 2 dimensions, to the topological $U(1)$ multiplet) is equivalent to a twisted version of a supersymmetric field theory in curved spaces. Therefore it is conceivable that study the PSMs coupled to the topological backgrounds (including topological gravity) could provide a way to define more general supersymmetric vector multiplets in 2 dimensions; in which the gauge group is replaced by, for example, a Lie algebroid.

Let us recall the field content and the BRST transformations of topological gravity \cite{Baulieu:1988xs}. The field content includes the 2-dimensional metric $g_{\mu\nu}$, the gravitino field $\psi_{\mu\nu}$, the diffeomorphism ghost $\xi^\mu$ and the ghost-for-ghost $\gamma^\mu$ which ensures the nilpotency of the BRST transformations. Such fields carry ghost numbers $0,\,1,\,1,\,2$ respectively and they transform as
\bea
\label{eq:topologicalgravitytransformations}
&& s \, g_{\mu\nu} = - \mathcal{L}_\xi g_{\mu\nu} + \psi_{\mu\nu} \ , \nn \\
&& s \, \xi^\mu = - \frac 12 \mathcal{L}_\xi \xi^\mu + \gamma^\mu \ , \nn \\
&& s \, \psi_{\mu\nu} = - \mathcal{L}_\xi \psi_{\mu\nu} + \mathcal{L}_\gamma g_{\mu\nu} \ , \nn \\
&& s \, \gamma^\mu = - \mathcal{L}_\xi \gamma^\mu \ ,
\eea
where $\mathcal{L}_\xi$ and $\mathcal{L}_\gamma$ are the Lie derivatives along the vector field $\xi^\mu$ and $\gamma^\mu$, respectively. Let us introduce the operator $S$
\bea
S \equiv s + \mathcal{L}_\xi \ ,
\eea
whose defining property is to satisfy, on all the fields but $\xi^\mu$, the relation
\bea
S^2 = \mathcal{L}_\gamma \ .
\eea
It is known \cite{Imbimbo:recent} that coupling a certain topological {\it matter} theory to topological gravity is equivalent to find a {\it new} BRST operator $S$, acting on the matter fields and satisfying also on them the relation
\bea
&& S^2 = \mathcal{L}_\gamma \ .
\eea
The general solution to this problem, for a theory formulated in terms of polyforms like our PSM, has been given in a 3-dimensional context in \cite{Imbimbo:2009dy}: it is sufficient to replace the coboundary operator $\delta_0$ introduced in (\ref{eq:coboundarydelta0}) with a new nilpotent operator $\delta$:
\bea
\label{eq:coboundarydelta}
 \delta \equiv S + \rmd - i_\gamma \  , \qquad \delta^2 =  0 \ . 
\eea

Given (\ref{eq:coboundarydelta}), we derive in a straightforward way the BRST variations for the fields and backgrounds of the PSM: in polyform notation they are
\bea
&& \delta \, \mathbf{f} = 0 \ , \nn \\
&& \delta \, \widetilde X^i = - \Pi^{ij}(\widetilde X) \, \widetilde \eta_j \ , \nn \\
&& \delta \, \widetilde \eta_i = - \frac 12 \, \partial_i \Pi^{jk} (\widetilde X)\, \widetilde \eta_j \widetilde \eta_k + \mathbf{f} \, \partial_i C(\widetilde X) \ ,
\eea
and in components they read
\bea
\label{eq:BRSTPSMcoupledgravityfinal}
&& S \, \beta_i = - \frac 12 \partial_i \Pi^{jk} (X)\, \beta_j \beta_k + \gamma^{(0)} \partial_i C(X) + i_\gamma (\eta_i)  \ , \nn \\
&& S  \, \eta_i = - \rmd \beta_i - \partial_i \Pi^{kl} (X) \,\eta_k \beta_l - \frac 12 \partial_i \partial_j \Pi^{kl} (X) \, \eta^{\ast j} \beta_k \beta_l + \nn \\
&& \qquad \quad + \partial_i C(X) \, \psi^{(1)} + \gamma^{(0)} \partial_i \partial_j C(X) \, \eta^{\ast j} + i_\gamma (X^\ast_i) \ , \nn \\
&& S \, X^{\ast}_i = - \rmd \eta_i - \partial_i \Pi^{kl}(X) \, X^{\ast}_k \beta_l - \partial_i\partial_j \Pi^{kl}(X)\, \eta^{\ast j} \wedge \eta_k \beta_l - \frac 12 \partial_i\Pi^{kl}(X)\, \eta_k \wedge \eta_l + \nn \\
&& \qquad  \; \; \; \; \; - \frac 14 \partial_i \partial_j \partial_p \Pi^{kl} (X) \, \eta^{\ast j} \wedge \eta^{\ast p} \beta_k \beta_l - \frac 12 \partial_i \partial_j \Pi^{kl} (X) \, \beta^{\ast j} \beta_k \beta_l  + \nn \\
&& \qquad  \; \; \; \; \; + f^{(2)} \, \partial_i C(X) + \partial_i \partial_k C(X) \, \bigl( \gamma^{(0)} \beta^{\ast k} + \eta^{\ast k} \wedge \psi^{(1)} \bigr) + \frac 12 \, \gamma^{(0)} \partial_i \partial_k \partial_j C(X) \, \eta^{\ast k} \wedge \eta^{\ast j } \ , \nn \\
&& S \, X^i =  - \Pi^{ij} (X) \, \beta_j + i_\gamma(\eta^{\ast i})\ , \nn \\
&& S \, \eta^{\ast i} = - \rmd X^i - \Pi^{ij} (X) \, \eta_j - \partial_k \Pi^{ij}(X) \, \eta^{\ast k} \beta_j + i_\gamma (\beta^{\ast i}) \ , \nn \\
&& S \, \beta^{\ast i} = - \rmd \eta^{\ast i} - \Pi^{ij}(X)\, X^{\ast}_j - \frac 12 \partial_k \partial_l \Pi^{ij}(X)\, \eta^{\ast k} \wedge \eta^{\ast l} \beta_j  + \nn \\
&& \qquad  \; \; \; \; \, - \partial_k \Pi^{ij} (X) \, \eta^{\ast k} \wedge \eta_j - \partial_k \Pi^{ij} (X) \, \beta^{\ast k} \beta_ j \ , \nn \\
&& S\, f^{(2)} = - \rmd\, \psi^{(1)} \ ,\nonumber\\
&& S\, \psi^{(1)} = - \rmd\, \gamma^{(0)} + i_\gamma (f^{(2)}) \ ,\nonumber\\
&& S\, \gamma^{(0)}= i_\gamma (\psi^{(1)}) \ .
\eea
Therefore we see that $S$ can be decomposed as
\bea
S \equiv s_0 + G_\gamma \ , 
\eea
where the nilpotent operator $G_\gamma$ is
\bea
\label{eq:Ggammaaction}
&& G_\gamma \, \beta_i = i_\gamma (\eta_i) \ , \qquad \qquad G_\gamma \, \eta_i = i_\gamma (X^\ast_i) \ , \qquad \quad G_\gamma \, X^{\ast}_i = 0 \ , \nn \\
&& G_\gamma \, X^i = i_\gamma (\eta^{\ast i}) \ , \qquad \quad \, G_\gamma \, \eta^{\ast i} = i_\gamma (\beta^{\ast i}) \ , \qquad \; \; \, G_\gamma \,\,  \beta^{\ast i} = 0 \ , \nn \\
&& G_\gamma \, \gamma^{(0)} = i_\gamma (\psi^{(1)}) \ , \qquad \; G_\gamma \, \psi^{(1)} = i_\gamma (f^{(2)}) \ , \qquad G_\gamma f^{(2)} = 0 \ ,
\eea
and $s_0$ is the {\it rigid} BRST operator (\ref{eq:BRSTPSMcoupledfinal}). The operators $s_0$ and $G_\gamma$ satisfy the $N = 2$ twisted supersymmetry relations
\bea
&& s^2_0 = G_\gamma^2 = 0 \ , \qquad \left\{ s_0 \, , \, G_\gamma \right\} = \mathcal{L}_\gamma \ .
\eea

Before concluding this Section, let us recall \cite{Bae:2015eoa} that one can treat both the topological gravity multiplet and the topological $U(1)$ multiplet $\mathbf f$ as external {\it rigid} backgrounds. In this way one looks for {\it bosonic} configurations of the backgrounds which are invariant under the action of the BRST operator $S$. The equations for such configurations read
\bea
\label{eq:invariantbackgroundsgravity}
&& \mathcal{L}_\gamma g_{\mu\nu} = 0 \ , \nn \\
&& \rmd \, \gamma^{(0)} = i_\gamma (f^{(2)}) \ .
\eea
The solutions of the system (\ref{eq:invariantbackgroundsgravity}) have been fully classified in \cite{Bae:2015eoa} where it has been also shown that they go in one-to-one correspondence with the bosonic supersymmetric solutions of $N = (2,2)$ supergravity, for which some explicit solutions were already known \cite{Closset:2014pda}.

\subsection{The observables and the action}
\label{sec:observablesactiongravity}

In this Section we will see that the observables introduced in Section \ref{sec:Observablescoupled} for the rigid theory are observables also for the theory coupled to topological gravity. We will then discuss how to modify the topological action (\ref{eq:PSMcoupledtwo}) to be invariant under the BRST operator coupled to topological gravity (\ref{eq:BRSTPSMcoupledgravityfinal}).

In the context of topological field theories coupled to topological gravity, the relevant cohomology is the {\it equivariant} one:  such cohomology is equivalent to the absolute cohomology of the coboundary operator $\delta$, defined in (\ref{eq:coboundarydelta}), on the space of polyforms which do {\it not} contain the diffeomorphisms ghost $\xi^{\mu}$.

To find representatives of the absolute cohomology  of $\delta$, let us observe that we have
\bea
\label{eq:deltadecomposition}
\delta \equiv S + \rmd - i_\gamma = s_0 + \rmd + G_\gamma - i_\gamma = \delta_0 + \delta_\gamma \ ,
\eea
where $\delta_\gamma$ is the nilpotent operator
\bea
\delta_\gamma \equiv G_\gamma - i_\gamma \ .
\eea

From (\ref{eq:deltadecomposition}) we see that the observables of the {\it rigid} model, i.e. the cocycles of the rigid coboundary operator $\delta_0$ are promoted to cocycles of the operator $\delta$ if they also satisfy the ``chirality'' constraint
\bea
\label{eq:chiralityGgamma}
&& \delta_\gamma \, \mathcal{O}^p \equiv (G_\gamma - i_\gamma) \, \mathcal{O}^p = 0 \ .
\eea
Let us note that the constraint (\ref{eq:chiralityGgamma}) is satisfied by all the superfields
\bea
&& \delta_\gamma \, \mathbf f = \delta_\gamma \, \widetilde X^i = \delta_\gamma \, \widetilde \eta_i = 0 \ ,
\eea
and therefore we conclude that the observables of the rigid model we discussed in Section \ref{sec:Observablescoupled}, formulas (\ref{eq:observablesrigidpolyform}) and (\ref{eq:dressedobservables}) are automatically promoted to observables of the model coupled to topological gravity, i.e. they satisfy the equation
\bea
\delta \, \mathcal{O}^p = 0 \ ,
\eea
(and of course an identical equation for $ \mathcal{O}^{p + 2 n}_{\mathbf f^n}$ ) which in components reads
\bea
&& S\, \mathcal{O}^{(2)}_p = - \rmd \, \mathcal{O}^{(1)}_{p + 1} \ , \nn \\
&& S\, \mathcal{O}^{(1)}_{p+1} = - \rmd \, \mathcal{O}^{(0)}_{p + 2} + i_\gamma (\mathcal{O}^{(2)}_p) \ , \nn \\
&& S \, \mathcal{O}^{(0)}_{p + 2} = i_\gamma( \mathcal{O}^{(1)}_{p+1}) \ .
\eea

Let us now discuss how to modify the topological action (\ref{eq:PSMcoupledtwo}) to mantain the invariance under the BRST operator $S$ (\ref{eq:BRSTPSMcoupledgravityfinal}). Looking at (\ref{eq:BRSTPSMcoupledgravityfinal}) we notice that the BRST variations for $\eta_i$ and $X^i$ have been modified by terms involving the antifields (or, in our framework, auxiliary fields) $X^\ast_i$ and $\eta^{\ast i}$. It is therefore natural to guess that the final action, to be invariant under (\ref{eq:BRSTPSMcoupledgravityfinal}), should contain additional terms which are {\it quadratic} in the antifields. Indeed, it can be verified by direct computation that the $\gamma^\mu$-dependent action
\bea
\label{eq:PSmcoupledtogravityaction}
&& \Gamma^C_{\rm PSM + \mathrm{grav} } [f^{(2)}, \gamma^{(0)} ,  C (X), \gamma^\mu] = \int_{\Sigma} \bigl[ \eta_i \wedge \rmd \, X^i + \frac 12 \, \Pi^{ij}(X) \, \eta_i \wedge \eta_j + \nn \\
&& \qquad \qquad - \frac 14 \eta^{\ast i} \wedge \eta^{\ast j} \, \partial_i \partial_j \Pi^{kl} (X) \,\beta_k \beta_l  - f^{(2)} \, C(X) + \nn \\
&& \qquad \qquad + \frac 12 \gamma^{(0)} \, \eta^{\ast i} \wedge \eta^{\ast j} \, \partial_i \partial_j C(X) + \eta^{\ast i} \wedge i_\gamma(X^\ast_i) \bigr] \ ,
\eea
is BRST invariant
\bea
S \,  \Gamma^C_{\rm PSM + \mathrm{grav} }  = 0 \ .
\eea

As we did in Section \ref{sec:actionPSMcoupledpolyform} we can construct a polyform of total ghost number $2$ in which the action (\ref{eq:PSmcoupledtogravityaction}) is the two-form representative
\bea
\label{eq:actioncoupledgravitypolyformone}
\mathcal{O}^{(2) g}_0 \equiv \Gamma^C_{\rm PSM + \mathrm{grav} }  \ .
\eea
By simply applying the BRST operator $S$ we obtain the one-form and the zero-form representatives
\bea
\label{eq:actioncoupledgravitypolyformtwo}
&& \mathcal{O}^{(1) g}_1 = \beta_i \rmd \, X^i - \psi^{(1)} C(X) + \eta_i i_\gamma(\eta^{\ast i})   \ , \nn \\
&& \mathcal{O}^{(0) g}_2 = - \frac 12 \Pi^{ij} (X) \, \beta_i \beta_j \ \gamma^{(0)} C(X) + \beta_i i_\gamma (\eta^{\ast i}) \ , 
\eea
which generalizes the anlogous expressions (\ref{eq:descentPSMcoupledaction}) which we discussed in the case of the rigid theory. It is interesting to note that the additional $\gamma^\mu$-dependent terms appearing in (\ref{eq:actioncoupledgravitypolyformone}) and (\ref{eq:actioncoupledgravitypolyformtwo}) allow to derive the relation
\bea
\label{eq:relationPSMcoupledpolyformgrav}
&& \mathcal{O}^{2 g} = \widetilde{\mathcal{O}}^2 + \delta\, (\alpha^\prime) \ , \nn \\
&& \alpha^\prime = \eta_i \wedge \eta^{\ast i} + \beta_i \beta^{\ast i} +\beta_i \eta^{\ast i} \ ,
\eea
between the polyform
\bea
 \mathcal{O}^{2 g} \equiv \mathcal{O}^{(2) g}_0 + \mathcal{O}^{(1) g}_1 + \mathcal{O}^{(0) g}_2 \ ,
\eea
and the algebraic polyform $\widetilde{\mathcal{O}}^2$ we defined in (\ref{eq:polyformactioncoupledalgebraic}). Notice that in (\ref{eq:relationPSMcoupledpolyformgrav}) it appears the coboundary operator $\delta$ and {\it not} its rigid counterpart $\delta_0$. Therefore we conclude that, also in the theory coupled to topological gravity, the topological  action (\ref{eq:PSmcoupledtogravityaction}) and the algebraic observable (\ref{eq:polyformactioncoupledalgebraic}) are in the same cohomology class.

\subsection{An application: the gauge-fixed vectorial supersymmetry}
\label{sec:gaugefixedsusy}

Let us describe a simple application of the coupling to topological gravity we constructed: we will see that the gauge-fixed vectorial supersymmetry discovered in \cite{Cattaneo:1999fm} has a transparent origin in the context of the theory coupled to topological gravity.\footnote{Our discussion in this Section can be considered as a simple rephrasing of the ideas developed in \cite{Imbimbo:2009dy} in a 3-dimensional context. Notice also that in this Section, in accordance with \cite{Cattaneo:1999fm}, we will adopt the BV point of view: the antifields will be considered as funtionals of the fields after the gauge fixing and the canonical piece will be included in the action.}

In \cite{Cattaneo:1999fm} the authors have noticed that, when the standard PSM is put on flat space\footnote{To be more precise, they considered the PSM on the disk. However, for what we are going to say, the presence of the boundary is not very important and can be neglected.} and the gauge-fixing fermion is taken to be
\bea
\label{eq:gaugefixingcattaneo}
\Psi = - \int \rmd \, \chi^i \, \ast \eta_i , 
\eea
where $\chi^i$ is the antighost forming a BRST doublet with the corresponding Lagrange multiplier $\lambda^i$
\bea
\label{eq:Ggammaauxiliary}
&& s_0 \, \chi^i = \lambda^i \ , \qquad s_0 \, \lambda^i = 0 \ , \nn \\
&& G_\gamma \, \chi^i = 0 \ , \qquad G_\gamma \, \lambda^i = \mathcal{L}_\gamma \chi^i \ ,
\eea
the theory develops a gauge-fixed vectorial supersymmetry  which reads
\bea
\label{eq:gaugefixedsusy}
\delta_\gamma \, X^i = i_\gamma (\ast \rmd \, \chi^i) \ , \qquad \delta_\gamma \, \beta_i = i_\gamma (\eta_i) \ , \qquad \delta_\gamma \, \eta_i =  \delta_\gamma \, \chi^i = 0 \ , \qquad \delta_\gamma \, \lambda^i = \mathcal{L}_\gamma \chi^i \ ,  
\eea
where $\gamma^\mu$ is a Killing vector. It is immediate to see that, given the gauge-fixing fermion $\Psi$ (\ref{eq:gaugefixingcattaneo}) and recalling the BV rules to identify the antifields with the derivatives with respects to the fields of the gauge-fixing fermion, one gets
\bea
\label{eq:gaugefixedantifields}
X^{\ast}_i = \beta^{\ast }_i = \lambda^{\ast }_i = 0 \ , \qquad \eta^{\ast i} = \ast \rmd \, \chi^i \ , \qquad \chi^{\ast}_i = \rmd \, (\ast \eta_i) .
\eea
Therefore we see that the transformations (\ref{eq:gaugefixedsusy}) are just the gauge-fixed version of the $G_\gamma$ transformations (\ref{eq:Ggammaaction}) and (\ref{eq:Ggammaauxiliary}).

However, we noticed in Section \ref{sec:observablesactiongravity} that the topological action of the standard PSM (\ref{eq:PSMactionone}) is in general {\it not} invariant under the $G_\gamma$ symmetry (\ref{eq:Ggammaaction}) and it must be completed by some $\gamma^\mu$-dependent terms. Including also the terms coming from the canonical piece (that in the BV formalism of \cite{Cattaneo:1999fm} must be added to the action), the $\gamma^\mu$-dependent terms read
\bea
\label{eq:gammadependentterms}
&& \eta^{\ast i} \wedge i_\gamma(X^\ast_i) \ , \qquad \beta^{\ast i} \, i_\gamma(\eta_i) \ , \qquad X^{\ast }_i \,  i_\gamma(\eta^{\ast i}) \ , \qquad  \lambda^{\ast }_i \, \mathcal{L}_\gamma \chi^i \ .
\eea
It is immediate to see that, in the particular gauge-fixing (\ref{eq:gaugefixingcattaneo}), all the $\gamma^\mu$-dependent terms appearing in (\ref{eq:gammadependentterms}) vanish thanks to (\ref{eq:gaugefixedantifields}).

Hence we conclude that the gauge-fixed vectorial supersymmetry found in \cite{Cattaneo:1999fm} is just the remnant, in flat space and in the particular gauge-fixing (\ref{eq:gaugefixingcattaneo}), of the $G_\gamma$-symmetry (\ref{eq:Ggammaaction}) and (\ref{eq:Ggammaauxiliary}). The vectorial supersymmetry is understood in \cite{Cattaneo:1999fm} as a property of the gauge-fixed action because only in the particular gauge-fixing (\ref{eq:gaugefixedantifields}) the $\gamma^\mu$-dependent terms in the topological action,  that would be necessary for the full $G_\gamma$-invariance and that in \cite{Cattaneo:1999fm} have not been considered, vanish.

\section{Conclusions and outlook}
\label{sec:conclusions}

In this paper we have discussed how the coupling of the 2-dimensional BF-model  to topological backgrounds \cite{Bae:2015eoa} can be generalized to an arbitrary 2-dimensional PSM (of which the 2-dimensional BF-model is just a particular case). Along the way we  have also reviewed a new application of the BV algorithm, application that has been  inspired by the study of supersymmetric field theories on curved spaces  and that has been developed for the first time in \cite{Imbimbo:2014pla} and \cite{{Bae:2015eoa}}: according to this supersymmetric point of view, the antifields are treated as independent {\it auxiliary} fields whose role is just to ensure the closure of the BRST algebra. It has been shown in \cite{Imbimbo:2014pla} and \cite{Bae:2015eoa} that this approach creates a natural connection between topological field theories and supersymmetric field theories on curved spaces. In the particular cases discussed in \cite{Bae:2015eoa} (and similarly in \cite{Imbimbo:2014pla}), one ends up with a correspondence between the topological system (in which both the topological $U(1)$ multiplet and 2-dimensional topological gravity are treated as {\it rigid} backgrounds and not as dynamical fields) and the vector multiplet in the corresponding  2-dimensional supersymmetric field theory. Given this connection, it is natural to conjecture that the study of the PSMs coupled to the topological backgrounds can be related to new supersymmetric vector multiplets, in which the gauge group is replaced by more general geometrical objects enjoying a Poisson structure, like for example a Lie algebroid. 

It would be very interesting to compute, via localization, some relevant quantities (i.e. the partition function and correlators of some observables) in the non-trivial topological backgrounds that solve (\ref{eq:invariantbackgroundsgravity}): for example, it is well-known \cite{Cattaneo:1999fm} that the correlators of the standard PSM sigma model on the disk reproduce the Kontsevich formula for the deformation quantization of the target space \cite{Kontsevich:1997vb}. It would be interesting to understand how the Kontsevich formula is modified by turning on the topological backgrounds.\footnote{It could be worth to recall that another, completely different, application of the 2-dimensional PSMs is in the description of the so-called dilaton gravity in 2 dimensions (See\cite{Grumiller:2002nm} for an exhaustive introduction). It would be interesting to understand if some of the results of this paper can be relevant in this direction.}

The coupling to the topological $U(1)$ multiplet makes use of a Casimir function $C(X)$ on the target space $M$. The resulting coupled system depends on the chosen Casimir function both in the BRST transformations rules (\ref{eq:BRSTPSMcoupledfinal}) and in the topological action (\ref{eq:PSMcoupledtwo}). Since the BRST transformation rules have been modified, the resulting system has different observables: before the coupling, the observables are in correspondence with the Poisson cohomology of the target space $M$ \cite{Bonechi:2007ar}; after the coupling  the observables are in correspondence with the Poisson cohomology elements which also commute with the Casimir function $C(X)$. We have also shown that these observables are promoted to observables of the theory coupled to topological gravity and that the topological PSM action is equivalent, in the relevant cohomology, to a purely algebraic observable (formulas (\ref{eq:relationPSMcoupledpolyform}) and (\ref{eq:relationPSMcoupledpolyformgrav})). 	

The coupling to topological gravity we described in Section \ref{sec:couplinggravity} is the first step to construct a topological string in propagation on the Poisson manifold $M$: on this aspect it is worth to recall that in the particular case in which $M$ is symplectic the PSM turns out to be equivalent to the topological A-model \cite{Bonechi:2007ar}, \cite{Bonechi:2016wqz}. Moreover, the A-model coupled to topological gravity computes the Gromov-Witten invariants of the target space. Given this observation it is tantalizing to conjecture that the coupling of the PSM to topological gravity gives a model that computes similar enumerative invariants for a generic Poisson manifold. For example, one could try to start with two particular examples: the case of the dual of a Lie algebra (i.e. consider 2-dimensional Yang-Mills coupled to dynamical topological gravity) and the case of a log symplectic manifold (see, for example, \cite{gualtierililogsymplectic} for the definition of such manifolds). Let us also notice that the construction provided in \cite{Imbimbo:2009dy} for the coupling of a rigid topological field theory to topological gravity is very general and can be applied all the times the rigid theory can be arranged in superfields: most of the TFTs of AKSZ type can be indeed arranged in superfields, and therefore we think that all that theories can be coupled to topological gravity in the same way as we did for the PSM in this paper.

Another, related, aspect that would be worth further investigation is the connection between the topological string theory we started to construct in this paper and some new integrable hierarchies, which should generalize the well-known correspondence between 2-dimensional topological gravity and the KdV hierarchy \cite{Witten:1990hr}.

Finally, it would be important to understand in full generality the relation between the BV formalism, in which the antifields are finally treated as functionals of the fields and they are used to gauge fix the local gauge symmetry, and the supersymmetric point of view we described in this paper and that goes back to\cite{Imbimbo:2014pla} and \cite{Bae:2015eoa}, in which instead the antifields are treated as independent fields and the new theory is in correspondence with a supersymmetric field theory on curved spaces: in the cases treated in \cite{Imbimbo:2014pla} and \cite{Bae:2015eoa}  the two approaches were trivially equivalent since the dependence of the action on the terms involving the antifields was quadratic and the local symmetry content of the theory was unchanged. On the other hand,  for a generic PSM we have seen that the situation is different and, for example, the local symmetry content gets changed when one treats the antifields as independent fields. Therefore we expect that the two approaches could be inequivalent in this case. 

We hope to address all these open questions in future works.

\section*{Aknowledgements}

I would like to thank Eduardo Conde, Noriaki Ikeda, Euihun Joung, Sungjay Lee,  Hai Lin, Dmitry Vassilevich  and, especially, Alberto S.~Cattaneo and Camillo Imbimbo for interesting discussions. I also thank the INFN Sezione di Genova and Camillo Imbimbo for the warm hospitality provided during part of this work.

\providecommand{\href}[2]{#2}

\end{document}